# Relationships between Software Architecture and Source Code in Practice: An Exploratory Survey and Interview

**Fangchao Tian [a,b], Peng Liang [a,*], Muhammad Ali Babar [b]**

[a] School of Computer Science, Wuhan University, 430072 Wuhan, China
[b] School of Computer Science, The University of Adelaide, SA 5005, Australia
tianfangchao@whu.edu.cn, liangp@whu.edu.cn, ali.babar@adelaide.edu.au

**Abstract**
**Context**: Software Architecture (SA) and Source Code (SC) are two intertwined artefacts that represent the interdependent design decisions made at different levels of abstractions - High-Level (HL) and Low-Level (LL). An understanding of the relationships between SA and SC is expected to bridge the gap between SA and SC for supporting maintenance and evolution of software systems.
**Objective**: We aimed at exploring practitioners' understanding about the relationships between SA and SC.
**Method**: We used a mixed-method that combines an online survey with 87 respondents and an interview with 8 participants to collect the views of practitioners from 37 countries about the relationships between SA and SC.
**Results**: Our results reveal that: practitioners mainly discuss five features of relationships between SA and SC; a few practitioners have adopted dedicated approaches and tools in the literature for identifying and analyzing the relationships between SA and SC despite recognizing the importance of such information for improving a system's quality attributes, especially maintainability and reliability. It is felt that cost and effort are the major impediments that prevent practitioners from identifying, analyzing, and using the relationships between SA and SC.
**Conclusions**: The results have empirically identified five features of relationships between SA and SC reported in the literature from the perspective of practitioners and a systematic framework to manage the five features of relationships should be developed with dedicated approaches and tools considering the cost and benefit of maintaining the relationships.

## 1 Introduction

Software Architecture (SA) represents the high-level structure of a software system. It characterizes the key design decisions taken to achieve the functional and non-functional requirements of a system [2]. From this perspective, SA is defined as a set of architecture design decisions [5]. Whilst SA has become a mature area of research and practice in Software Engineering (SE) over the last 30 years (since 1980s) [11], there is still no consensus on what exactly SA is [1]. The representation of SA is described from different perspectives by using architectural styles, views, patterns, tactics and decisions [5]. Due to the difference in the abstraction levels (i.e., HL and LL), there exists a gap between SA and Source Code (SC) [10]. This gap leads to a problematic situation exacerbated by a general lack of efforts allocated to keep the SA documentation updated [58]. Even if SA is well documented or updated, it is still a challenge about how to embody SA in implementation and update SA and SC synchronously for ensuring their consistency during software evolution [7]. Several different approaches already address the problem of narrowing the gap to align and evolve SA with SC over time [11]. During the software development life cycle, architectural level elements, e.g., modules, components, decisions, tactics and patterns are transformed to low-level design elements to be implemented [39]. SA elements of a system can also be reconstructed by reengineering the SC elements of a system [30][35]. When an SA is modified to satisfy new requirements, its corresponding SC also gets changed when a software system evolves for avoiding architecture erosion [7]. When anomalies in architecturally relevant SC are detected and corrected, the corresponding SA problems should also be fixed to maintain a system [8]. It is important to analyze and leverage the relationships between SA and SC to effectively improve the maintenance and evolution of a software system.

---

* Corresponding author at: School of Computer Science, Wuhan University, China. Tel.: +86 27 68776137; fax: +86 27 68776027. E-mail address: liangp@whu.edu.cn (P. Liang).

The relationships between SA and SC collected in this study include the links that exist between SA and SC (e.g., traceability relationship), the activities that involve SA and SC (e.g., recovering SA from code), and the properties that are satisfied between SA and SC (e.g., consistency between SA and SC). Researchers have empirically investigated certain types of relationships between SA and SC, such as the Systematic Literature Review (SLR) on traceability relationship between SA and SC [15], the empirical evidence in the literature about consistency between SA and SC [24], and the comparative analysis of architecture recovery techniques [34]. The SLR on architectural traceability [15] reveals that no empirical studies were found to explore the architectural traceability from the perspective of practitioners. The SLR on consistency between SA and SC [24] reveals that only a few industrial surveys were conducted to investigate such consistency. Although these empirical studies [21][29][31][32][59][60] intended to understand practitioners' perception and feedback about how they used current approaches and tools for architecture consistency and recovery, to our knowledge, there has been no empirical study conducted to comprehensively investigate the potential relationships between SA and SC from the practitioners' perspectives. Hence, there is an important need of exploring the SE practitioners' perceptions and practices of dealing with the relationships between SA and SC.

We decided to empirically study the practitioners' understanding of the relationships between SA and SC with the aim of bridging the gap between SA and SC of a software system. The objectives of this survey were to explore: (1) The state of the practice on how architects and developers identify, analyze, understand and use the relationships between SA and SC for bridging the gap between SA and SC. (2) The difficulties and benefits of identifying, analyzing and using the relationships between SA and SC during system evolution. (3) The difference in the views of industrial practitioners and academic researchers on the relationships between SA and SC.

We used an online survey and interview to explore the above objectives. We initially conducted an online survey by following the guidelines provided by Kitchenham and Pfleeger in [12]. We sent the survey questionnaire to approximately 1000 potential respondents, whose contact details were extracted from GitHub. We received 87 valid responses, which were analyzed using qualitative and quantitative data analysis approaches. Then the survey results were further used and validated by the interview with 8 practitioners. **The findings of this study are that**: (1) Practitioners mainly discuss five features of relationships between SA and SC; (2) Practitioners seldom used dedicated approaches and tools reported in the literature for establishing and understanding the relationships between SA and SC; (3) Systems qualities (e.g., maintainability and reliability) can be improved by identifying, analyzing, and leveraging the relationships between SA and SC. But project resourcing difficulties (e.g., cost and effort) discourage practitioners from fully understanding and leveraging the relationships between SA and SC. **The findings of this study show that**: (1) More attention of practitioners should be given to Interplay and Recovery relationships between SA and SC, as well as the dedicated tools proposed in the literature to manage these relationships. (2) Researchers should put more effort into how to systematically manage the five relationship features between SA and SC, and alleviate the cost and effort for bridging the gap between SA and SC.

The rest of the paper is structured as follows: The background is introduced in Section 2 followed by the related work in Section 3. The study design including the research objectives and questions, survey design, and interview design are described in Section 4. The results of the survey and interview are presented in Section 5, and the findings and implications are discussed in Section 6. The potential threats to validity are discussed in Section 7. This paper concludes in Section 8 by highlighting the future research directions.

## 2 Background

Software architecture is considered a key research and practice area of Software Engineering (SE) [11]. Medvidovic and Taylor highlighted the transition in developers' focus from Line-of-Code (LoC) to coarser-grained architectural elements (e.g., components, decisions, patterns, and styles) and their overall interconnecting structures (e.g., connectors) [3]. Fairbanks [10] pointed out the inherent difference between the way architects reason about systems in terms of components, layers, decisions at the architecture level and the way developers work with classes, packages, and interfaces at the implementation level. Hence, it is becoming important to software developers to build and understand the correct links between SA and SC artefacts. If there are no suitable links between the abstraction levels of SA and SC, there is a potential risk of treating SA and SC as two unrelated artefacts in software development [9]. That is why Fairbanks calls this difference the "model-code gap" [10]. The intersection between SA and SC has been the focus of research to bridge the gap between SC and high-level models [28]. Recently, a lot of research efforts have been devoted to this focus, by construction (via automatic code generation and formal refinement) or extraction (analyzing the implementation statically or dynamically to determine its architecture) [11].

## 3 Related Work

We discuss how the existing approaches have been adopted to build and maintain the relationships between SA and SC for bridging the gap. We also summarize the literature reviews and the practitioners' studies on this topic.

### 3.1 Existing Approaches and Tools

Some studies focus on automatic code generation from SA models. A model-driven approach of automatic code generation has been reported in [37]. In this approach, a reference architecture and a domain-specific language are predefined to transfer architectural design to implementation. Other research supports code generation from Architecture Description Languages (ADLs), such as AADL, xADL, and PRISMA [40]; for example, an automatic C code generator was designed for code generation from AADL to bridge the gap between SA and SC [39].

Another set of studies have focused on building and analyzing trace links between SA and SC. Buchgeher and Weinreich presented a semi-automatic approach to capture the traces from decisions to architecture and implementation as the basis for architecture-related activities [16]. Another approach was proposed by Nguyen *et al.* with an architectural Software Configuration Management (SCM) system, MolhadoArch, to make configurations at the architecture and implementation levels uniform by managing the traceability relationships between SA and SC [18]. A web-based tool, BUDGET, was developed by Santos *et al.* to support architecture traceability by automatically generating training data of architectural tactics related to code snippets [19]. Another tool, ArchTrace, was developed to continuously update and manage traceability relations between SA and SC through a policy based extensible infrastructure [7]. Zheng and Taylor proposed 1.x-way mapping approach and an Eclipse-based tool, xMapper, that supports the behavioral architecture-implementation mapping [20].

Different methods have been proposed to extract architecturally relevant artefacts from implementation [30]. The extracted artefacts can help to reconstruct SA that can represent a system's requirements [33]. Mirakhorli and Cleland-Huang proposed a Machine Learning (ML) based approach to automatically detect architectural tactics from SC for building the links between architectural tactics and SC [36]. Shahbazian *et al.* developed RecoAr to automatically recover architectural decisions from an issue tracker and version control repository [35].

Architecture recovery has been employed as a technique to ensure architecture consistency, which is also known as architecture conformance or compliance. This is an important measure to deal with architecture erosion or drift during a software system' evolution [22][23][24]. Passos *et al*. reviewed three static architecture consistency checking techniques, i.e., Dependency Structure Matrix (DSM), source code query languages, and reflexion models [27], but none of these techniques can check architectural consistency based on dynamic information, such as the execution of method calls. Several approaches and tools can check architecture consistency by using dynamic analysis on an executing system. For example, Muccini *et al.* proposed SA-based code testing approaches that use SA as a reference model for code conformance testing, which can dynamically verify whether a program's behaviors conform to its architecture behaviors [25][26]. These SA-based testing approaches are supported by a mapping between SA-based and code-based test cases. Rosik *et al.* developed a tool, ACTool, which embodies real-time RM for architecture recovery and compares the consistency between recovered architecture and planed architecture [30]. De Silva and Balasubramaniam [62] proposed PANDArch as a dynamic architecture conformance checking framework by using Java Virtual Machine to capture the mapping between the runtime system and the system architecture. In another example, Ganesan *et al*. [63] constructed hierarchical colored Petri nets to extract architectural views from runtime traces for checking architecture compliance.

Based on this real-time RM, Buckley and his colleagues [22] also reported an Eclipse plug-in, JITTAC, to define the mapping between SC and architectural models, which can support static architectural consistency that architectural models are changed instantaneously when code is updated to detect and remove architectural drift. Stevanetic *et al.* proposed a Domain Specific Language (DSL)-based architecture abstraction approach to automatically generates traceability links and enables consistency checks between an architectural component view defined with DSL and the underlying SC [17]. According to Woods and Rozanski, a lack of architectural information in implementation is the major obstacle to narrow the gap between SA and SC of a software system [41]. To address this problem, researchers have developed some approaches to integrate SA concepts into programming languages. Aldrich *et al.* proposed a Java-based architectural programming language, ArchJava, which integrates SA specifications seamlessly into Java implementation code for ensuring traceability and consistency between SA and SC [38]. Pelliccione *et al.* presented a framework, CHARMY, to automatically generate Java code conforming to structural SA constraints [40].

There are also several studies that focus on evaluating SA quality by analyzing code. Fontana *et al.* focused on assessing SA quality from the perspective of exploiting code smell relations [42]. Pigazzini proposed two detection strategies of concern-based architecture smells using the semantic representation of code [43]. Vidal *et al*. studied architecturally relevant code anomalies that some types of code smells agglomerations could well reflect SA problems. They presented five criteria for prioritizing groups of code smells as indicators of architectural problems [44]. Moreover, Santos *et al*. found that the flaws in the implementation of security tactics can introduce severe vulnerability and identified the most occurring vulnerability types related to architectural security tactics in three large-scale open source projects [45].

Whilst the abovementioned approaches have been proposed to build and maintain the relationships between SA and SC, none of them has been studied from the practitioners' perspectives. Hence, one of the key motivators of our research presented in this paper is to investigate whether or not the above discussed dedicated approaches and tools have been adopted by practitioners; we are also interested in exploring the difference between industry and academia on the approaches and tools for bridging the gap between SA and SC.

### 3.2 Literature Studies

It seems permanent to briefly mention some of the existing literature-based studies aimed at investigating the relationships between SA and SC. Javad and Zdun carried out an SLR to discover the state-of-the-art of traceability approaches and tools between SA and SC [15]. Another study presented an empirical evidence from the literature about AC checking and the results show that more surveys are required to improve the validity and evidence regarding cause and effect relationships related to AC [24]. Moreover, in a literature survey, Silva *et al.* reported a lack of widespread adoption of AC checking strategies in industry [23]. Though these studies have provided favorable results, they only report on the specific types of relationships between SA and SC from the existing literature. Our study is aimed to investigate different aspects of relationships between SA and SC by exploring the views and experiences of practitioners.

### 3.3 Practitioner Studies

These works (i.e., the SMS [15] on architectural traceability, the SLR [24] on architectural consistency, and the literature survey [61] on architectural recovery) reveal that few studies have investigated the needs of practitioners about architectural traceability, consistency, and recovery. Buckley *et al.* reported a multi-case study that evaluated the use of a tool, ACTool, to support architecture recovery and architecture consistency [32]. In their study, a multi-case study was conducted to assess the insights of professional architects about the capabilities of Real-Time Reflexion Modelling (RT-RM). Le Gear *et al.* [59] proposed a technique Software Reconnexion to recovery architecture. One practitioner was involved in a case study to evaluate how this technique can help him recover architectural elements and understand an unfamiliar system. Brunet *et al.* studied how developers of the open source Eclipse platform checked architecture consistency by analyzing five years of the Eclipse architecture-checking reports [31]. However, these studies [29][31][32][59] aimed to explore the effectiveness of particular architecture consistency tools used by developers in certain projects. Similar to our work, Ali *et al.* also reported an empirical investigation of the practitioners' perception of consistency between SA and SC by interviewing 19 senior software engineers [21]. However, their work is mainly focused on investigating the needs for architecture consistency approaches and tools or whether the existing architecture consistency tools meet the identified needs. Wohlrab *et al.* [60] reported two industrial surveys with 93 and 72 practitioners respectively about how architectural inconsistencies were fixed over time. Their survey results indicate that consistency between architecture and code is still a prevalent issue in practice. In contrast, our work investigates the understanding of practitioners about the potential relationships between SA and SC for narrowing down and bridging the gap between them.

## 4 Study Design

### 4.1 Objective and Research Questions

As previously stated, we aimed at empirically investigating the relationships between SA and SC from the perspective of practitioners. The objectives of this study are to: *identify the **practitioners' perceptions** of the relationships between SA and SC; identify the **methods and tools** used to identify, analyze, and use the relationships between SA and SC; investigate the **benefits and limitations** of identifying, analyzing, and using the relationships between SA and SC; compare the difference between the academia and industry about the views on the relationships between SA and SC.*

To achieve the abovementioned objectives, we formulated and investigated three Research Questions (RQs) enlisted in Table 1.

Table 1. Research questions and their rationale

| Research Question | Rationale |
|---|---|
| **RQ1**: What are the practitioners' perceptions of the relationships between SA and SC? | This RQ is aimed at obtaining the answers about understanding and classifying the relationships between SA and SC from the perspectives of practitioners, which can help practitioners to systematically manage diverse relationships between SA and SC. |
| **RQ2**: What are the approaches and tools used for identifying, analyzing and using the relationships between SA and SC?<br>  **RQ2.1**: What are the approaches used to identify, analyze and use the relationships between SA and SC?<br>  **RQ2.2**: What are the tools used to identify, analyze and use the relationships between SA and SC? | The answer of this RQ provides information about what specific approaches and tools are employed by practitioners to identify, analyze and use certain relationships between SA and SC; this information can also help researchers to identify the gap between academia and industry in the approaches and tools. |
| **RQ3**: What are the benefits and limitations of identifying, analyzing and using the relationships between SA and SC? | The answer of this RQ helps practitioners make a trade-off between the benefits and limitations of identifying, analyzing and using the relationships between SA and SC, and helps researchers be aware of the difficulties and challenges to be addressed in identifying, analyzing and using the relationships between SA and SC. |

Our study methodology follows a mixed approach including two stages (survey and interview) to collect data from practitioners (see Fig. 1). In the first stage, we conducted an online survey to initially investigate the practitioners' perception of the relationships between SA and SC. In the second stage, we designed an interview instrument to further explore and validate the initial findings from the online survey. In the following subsections, we present the detailed process of each stage.

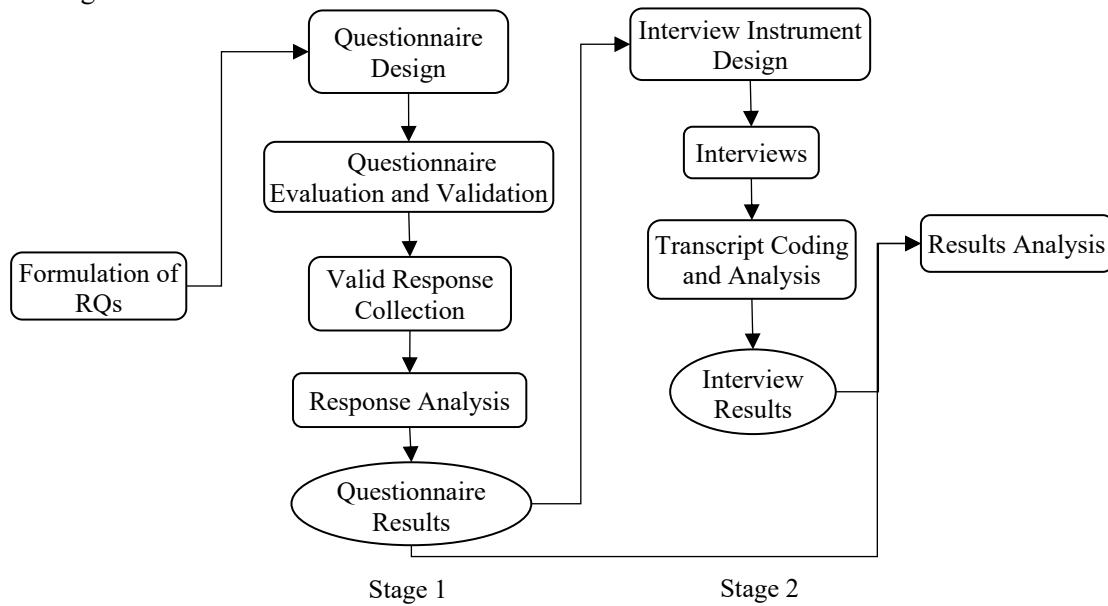

Fig. 1. The two stages of the study.

## 4.2 Survey Design

A survey is a "*system for collecting information from or about people to describe, compare, or explain their knowledge, attitudes and behavior*" [13]. Considering the research objectives and RQs, we decided to carry out a survey as a data collection tool to understand practitioners' perceptions and current practices in the relationships between SA and SC. Firstly, we chose a cross-sectional survey because we wanted to collect participants' experiences at a particular point in time. Based on the survey study guidelines [12], a survey research usually utilizes self-administered questionnaires, telephone surveys, and one-to-one interviews as instruments to gather data. The pros and cons of different instruments are discussed in detail [14]. Although several drawbacks of self-administered questionnaires, we employed a descriptive survey with an online self-administered questionnaire to collect information by considering the facts that we wanted to obtain the evidence from a wide variety of subjects who are distributed in diverse locations and willing to answer batteries of similar questions in an online questionnaire.

### 4.2.1 Creating the questionnaire

We formulated our study's questionnaire based on the pre-defined research questions and a thorough review of the relevant studies in Section 3. The questionnaire was formulated in English, as we planned to invite potential participants from multiple countries. We developed and reviewed the questionnaire consisting of both closed-ended and open-ended questions (e.g., Likert-Scale questions, Multiple-Choice questions and a free text option). As shown in Fig. 2, the questionnaire is composed of five parts: the welcome page, eight questions about the background information of the participants, six questions for answering RQ1, two questions for answering RQ2, two questions for answering RQ3. In total, we defined 18 questions for the full questionnaire (see Appendix A), which is also available online in [46]. The questionnaire was created using a survey administration platform - Google Forms[1], with which all the authors utilized real-time collaboration to evaluate and refine the questionnaire.

In the first part of the questionnaire, as shown in Table 12, we described the purpose and certain requirements for the potential participants of this study. We also explained that the questionnaire could be completed within 10-15 minutes in this part. 10-15 minutes are estimated completion time for answering the questionnaire according to the pilot result in Section 4.2.2, but the actual time spent by the participants cannot be measured due to the limitation of Google Forms. We explicitly stated that the personal information of the respondents will be kept confidential. These statements can create a level of trust between researchers and participants of a study.

For the second part (see Table 13), we asked the participants about their demographic information with specific characterization questions (Q1 to Q8) such as their education and experience. Among these questions, we have some questions (Q3 to Q5) to ask the participants about their roles and level of experience in software development. Such information was expected to help us to assess the participants' eligibility for the survey based on the inclusion and exclusion criteria defined in Table 2. The data from the demographic questions was also analyzed to understand the impact of the participants' backgrounds on the results. The last three parts in Table 13 are respectively corresponding to the three RQs. The detailed survey questions are provided to gather information for answering the RQs. In the third part, we designed six survey questions (Q9 to Q14) to identify the practitioners' perceptions of the relationships between SA and SC. Initially, we asked (Q9) whether the participants know certain relationships between SA and SC. Next, we asked the participants about the reasons (Q10) and the associated examples of the relationships between SA and SC (Q11). Then, we asked the participants whether they ever considered these relationships during their work (Q12) and to provide some examples (Q13). Finally, we asked the participants about the importance of identifying, analyzing and using the relationships in daily software development (Q14). In the fourth part, we designed two survey questions Q15 and Q16 in order to understand how practitioners identify, analyze and use the relationships between SA and SC with the support of the approaches and tools in their work. In the last part, we designed Q17 and Q18 to respectively investigate the benefits and limitations of identifying, analyzing and using the relationships between SA and SC.

---

[1] https://www.google.com/forms/about/

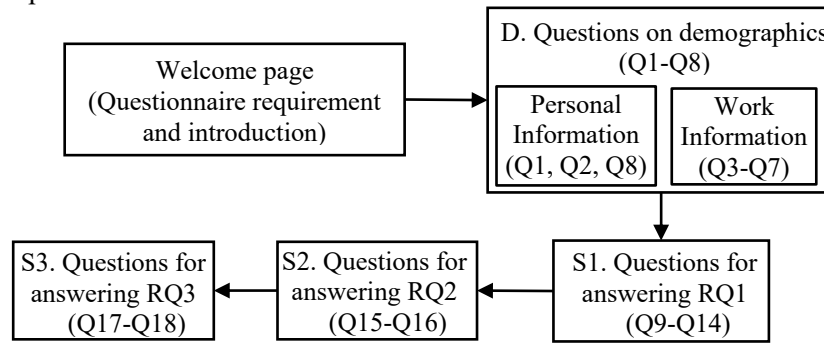

Fig. 2. Flow of the questionnaire

As observed in the questionnaire [46], 61% of the survey questions are open-ended questions (i.e., Q1, Q4, Q5, Q10, Q11, Q13, Q15, Q16, Q17, and Q18). Open-ended questions are expected to enable respondents to frame their own response and avoid restriction and bias on the responses of the participants to collect qualitative data. Moreover, open-ended questions can produce a much more diverse set of answers than close-ended questions, which are suitable for exploratory and descriptive survey with flexibility [4]. The Q2, Q3, and Q7 are semi-structured questions with several optional answers and a textual answer. The Q6, Q9, Q12, and Q14 are close-ended questions with a limited set of possible answers. These questions were constructed with a categorical question (Q6) and multiple-choice questions (i.e., Q9, Q12, and Q14). All the questions are mandatory with one exception that the Q8 is to ask the email address of the participants for sharing survey results because it can raise the curiosity and motivation of the participants of a study [12]; however, the Q8 was not a mandatory question in case the participants did not want to answer for privacy reasons. Without filling in all the required questions, the participants could not submit the questionnaire online, which might have been one reason that we got a low response rate (see Section 4.2.3).

### 4.2.2 Evaluating and validating the questionnaire

Before putting the survey questionnaire online for gathering data, the authors reviewed the protocol of this survey to reach a consensus on the survey instrument, such as survey objectives and survey questions. Furthermore, two other researchers in the SA domain gave feedback on the protocol of this survey. The authors also discussed and refined the protocol according to the feedback; for example, clarifying the rationale of RQs, refining the survey questions.

Once the survey was developed, a pilot survey with five practitioners with the relevant experience of architecture design was conducted to further evaluate the instrument's reliability [12]. By conducting the pilot study, we checked the understandability of the questions, the amount of time needed to answer the questions, and the response rate and the effectiveness of each of the questions. Two questions asked the respondents about the means through which they would identify and analyze the relationships between SA and SC. After analyzing the answers from the five respondents about these two questions, we found that some respondents could not differentiate between identifying and analyzing the relationships between SA and SC; consequently, we merged the two ambiguous survey questions into one question about how the participants identify and analyze the relationships between SA and SC. Based on the feedback from the pilot study, we prepared a final version of the questionnaire consisted of 18 questions that could be answered within 10-15 minutes.

#### 4.2.3 Collecting valid data

In this section, we describe how to define the target population for obtaining a valid sample with non-probabilistic sampling methods. We also analyzed the response rate of the participants after implementing the web-based survey with a selected sample size of the target population.

**Sampling and populations**. The target population is selected based on a survey objective. We intended to investigate the software practitioners' understanding of the relationships between SA and SC. We did not limit the domain and countries in which the participants were working. We defined our target population as the practitioners who have experience or get involved in software development and architecture design. It is very hard to approach and invite every member of the target population using a probabilistic sampling method, especially from unfamiliar sources. Considering the appropriateness and cost-effectiveness of recruiting the potential participants. We used non-probabilistic sampling methods, i.e., convenience sampling and snowball sampling, to reach the target population with multiple strategies. Firstly, we utilized convenience sampling to recruit our previous contacts in industry who were available and willing to participate in a survey study. We also used GitHub[2] to search for the potential participants with the keyword "architect" in users' profiles. We filtered out the retrieved users by checking their repositories in GitHub or personal websites provided in GitHub. We assessed the users on GitHub as eligible practitioners for the survey if they meet these criteria: (1) their affiliated organizations in industry were provided in their GitHub homepages or the profiles in the linked curricula vitae; and (2) their experience of architecture design and software implementation was described on the profiles in the linked curricula vitae. Moreover, we employed snowballing sampling and asked the participants of this survey to invite their social network contacts who might have been willing to take part in this industrial survey.

**Obtaining a valid sample**. After we reached the target population, we contacted the potential participants and sent them direct invitations with the questionnaire URL link [46] through emails or other social media (e.g., LinkedIn, Twitter, and Facebook) depending on the types of contacts they had provided in GitHub. We sent the questionnaire to 1000 potential participants by counting the number of emails and messages in social media we sent to individuals. We received 103 responses with an approximate response rate of 10.3%. We defined the inclusion and exclusion criteria in Table 2 to select the valid responses for data analysis. Among the 103 responses, ten invalid responses were excluded based on the criteria set for the respondents; these ten respondents had the required experience in architecture design and implementation but their answers were irrelevant. Another six responses were excluded because these respondents did not have experience in architecture design although they provided meaningful responses. Finally, we retained 87 valid responses (i.e., **R1** to **R87)** for data analysis.

Table 2. Inclusion and exclusion criteria to select valid responses

| Inclusion Criteria |
|---|
| I1: The respondent has experience in architecture design and code development. |
| I2: The respondent has English ability to answer the survey. |
| **Exclusion Criteria** |
| E1: The response is not in English. |
| E2: The response is meaningless, e.g., the response contained clear and significant inconsistencies. |

#### 4.2.4 Analyzing survey data

We used descriptive statistics [48] to quantitatively analyze the data gathered from the answers to the close-ended questions on the demographic information of the respondents. Moreover, we used systematic qualitative data analysis techniques [49] to analyze the answers to the open-ended questions. Table 3 shows the relationship between survey questions, data analysis methods, and RQs. Two steps (open coding and selective coding) of the classical Grounded Theory (GT) [50] were used to analyze the data gathered from the answers to certain survey questions (e.g., Q10, Q11, Q13, Q15, Q16, Q17, Q18). We used these two steps to codify the qualitative data, generate concepts and categories from the codes. A qualitative data analysis tool MAXQDA[3] was used to support the Grounded Theory analysis process. The first author read the answers to the open-ended questions and generated codes from the indicators (e.g., words, sentences, or paragraphs) provided by the respondents. Then, the generated codes were clustered into concepts. Moreover, subsequent indicators were constantly compared to the existing indicators and emerging concepts until no new concepts emerged. For example, we generated some codes, e.g., "Architecture guides the implementation of code", "Architecture guides development strategies", which were further merged into the concept of "Architecture leads or guides code", one subfeature of Transformability relationship (see Section 5.4.1). Secondly, the first author used selective coding to identify the core categories (e.g., the types of approaches in Table 8) that were generated by aggregating a set of concepts. Note that we only used open coding to analyze the answers to the survey question Q17 and Q18, since only concepts were generated from the answers. To reduce the personal bias during the data analysis, the other authors participated in the validation of the generated code. The disagreements were discussed and resolved. We finally generated 589 codes, 36 concepts, and 21 core categories from the gathered data. The survey results are reported in Section 5 and used to guide the design of the interview instrument. We also provided the survey responses in MS Excel and the encoded data in MAXQDA online [47].

---
[2] https://github.com/

Table 3. Relationship between survey questions, data analysis methods, and research questions

| Survey Question | Data Analysis Method | RQs |
| --- | --- | --- |
| Q1-Q8 | Descriptive Statistic | Demographics |
| Q9, Q12, | Descriptive Statistic | RQ1 |
| Q14 | Descriptive Statistic | RQ2 |
| Q10, Q11, Q13 | Open Coding, Selective Coding, and Constant Comparison | RQ1 |
| Q15, Q16 | Open Coding, Selective Coding, and Constant Comparison | RQ2 |
| Q17, Q18 | Open Coding and Constant Comparison | RQ3 |

## 4.3 Interview Design

We conducted an interview to further validate and confirm the initial findings from the online survey. In the following subsections, we describe the interview instrument, participants selection, and data analysis.

### 4.3.1 Interview instrument

We designed a semi-structured interview instrument containing 22 questions (IQ1-IQ22) that are formulated based on the findings of the online survey. The instrument is available in Appendix B. The instrument is composed of four parts: the welcome massage in Table 14, 14 questions (IQ2.1.1-IQ2.1.8, IQ2.2.1-IQ2.2.3, IQ2.3) for validating the findings of RQ1, two questions (IQ3.1-IQ3.2) for validating the findings of RQ2, two questions (IQ4.1-IQ4.2) for validating the findings of RQ3 (see Table 15). Since the interview is semi-structured, the list of questions in the instrument is only considered as a reference for the interviewers. In the interview process, we do not enforce interviewees to answer the open-ended questions in a controlled order so that they can naturally follow the conversation and provide their knowledge and experience of pertinent topics. Moreover, each interview is supposed to be conducted between 30 and 45 minutes according the pilot interview with one participant as detailed below. Note that the pilot interview result was included in the formal interview results since the interview instrument was not modified based on the pilot interview result.

---

[3] https://www.maxqda.com/

#### 4.3.2 Participants selection

The selection criteria for interview participants are the same as the criteria used for the online survey in Table 2. We used purposive sampling and convenience sampling to select the interview participants where the potential interviewees were first recruited from the survey respondents and then from our personal networks. Using this recruiting strategy, we acquired 8 practitioners for our interview. These 8 interviewees are referred as **IP1** to **IP8**. All the interviews were conducted remotely in English or Chinese by the first author using Zoom or Skype. The interview process was video-recorded with the permission of interviewees for further transcription and analysis. We firstly piloted the interview instrument with one architect (**IP1**) to verify the validity of the interview instrument. The result of the pilot interview shows that the interviewee can understand each question and complete the interview within 45 minutes. Moreover, the participants completed the interview with 37 minutes on average. Out of the eight interview participants, two (i.e., **IP1**, **IP3**) are from the survey respondents; the other six are newly recruited from our personal contact. The demographic data of interviewees are reported in Table 4. Note that the demographic information of **IP1** and **IP3** was asked again and updated.

#### 4.3.3 Data analysis

The recorded videos of interview were converted into verbatim transcripts by using the free transcription service provided by an online NetEase Sight platform[4]. We then proofread the transcripts to rectify the incorrect words. Note that, since two interviewees (i.e., **IP1** and **IP2**) participated in our interview in Chinese, videos of these two interviewees were initially converted into Chinese transcripts and then translated into English transcripts. Descriptive statistics [48] was used to analyze the quantitative data about interviewees demographics; open coding and selective coding of the classical Grounded Theory (GT) [49] were used to analyze the answers to open-ended questions. MAXQDA was also used to facilitate the process of qualitative data analysis. The data analysis was conducted by the first author, whereas the second and third authors verified the coding results and resolved the disagreements with the first author for ensuring the quality and credibility of coding results. We also provided the transcripts in MS Word and the coding results in MAXQDA online [47]. The interview results are used to validate and refute the results of the online survey in Section 5.

## 5 Results

We present the survey results by analyzing the respondents' demographics and answering the three RQs defined in Table 1. For supporting the clarity about the coding results, we use excerpts from the participants' responses as examples. The following four subsections are described corresponding to the four parts of the used questionnaire. Since the objective of conducting interviews is to validate and refute the results of the online survey, we discuss the results of the interviews across Section 5.2, Section 5.3, and Section 5.4 in conjunction with the survey results.

### 5.1 Demographic Data

Fig. 3 shows that 87 survey respondents are residing in 37 countries indicating that the international scope of the survey is very large. Nearly half of the respondents (49.4%, 43 out of 87) are located in countries with English as an official language. Moreover, the majority of them (44.2%, 19 out of 43) are from USA, followed by Pakistan, India, and the UK.

The educational level of the respondents is shown in Fig. 4(a). Most of the respondents (81.6%, 71 out of 87) have a B.Sc. degree or above, including a B.Sc. degree of 32 (36.8%, out of 87) respondents, an M.Sc. degree of 36 (41.4%, out of 87) respondents, and a PhD degree of 3 (3.4%, out of 87) respondents. Among the respondents without a formal degree, seven of them are self-taught and four respondents are university dropouts.

---

[4] https://sight.youdao.com/

In terms of the major roles in their projects, the respondents can select multiple responsibilities. As shown in Fig. 4(b), a large majority of them (85.1%, 74 out of 87) work mainly as an architect, followed by the role of a developer (72.4%, 63 out 87). We observed in Fig. 5 that all respondents are experienced in both architecture and implementation. More specifically, 78.2% (68 out of 87) of the respondents have more than 5 years of experience in software development with 46.0% (40 out of 87) having more than 10 years of experience (see Fig. 5(a)). In terms of software architecture experience shown in Fig. 5(b), 58.6% (51 out of 87) of the respondents have more than 5 years of experience with 31% (27 out of 87) having more than 10 years of experience. These evidences give us confidence that our respondents have the eligible experience to complete the survey questionnaire.

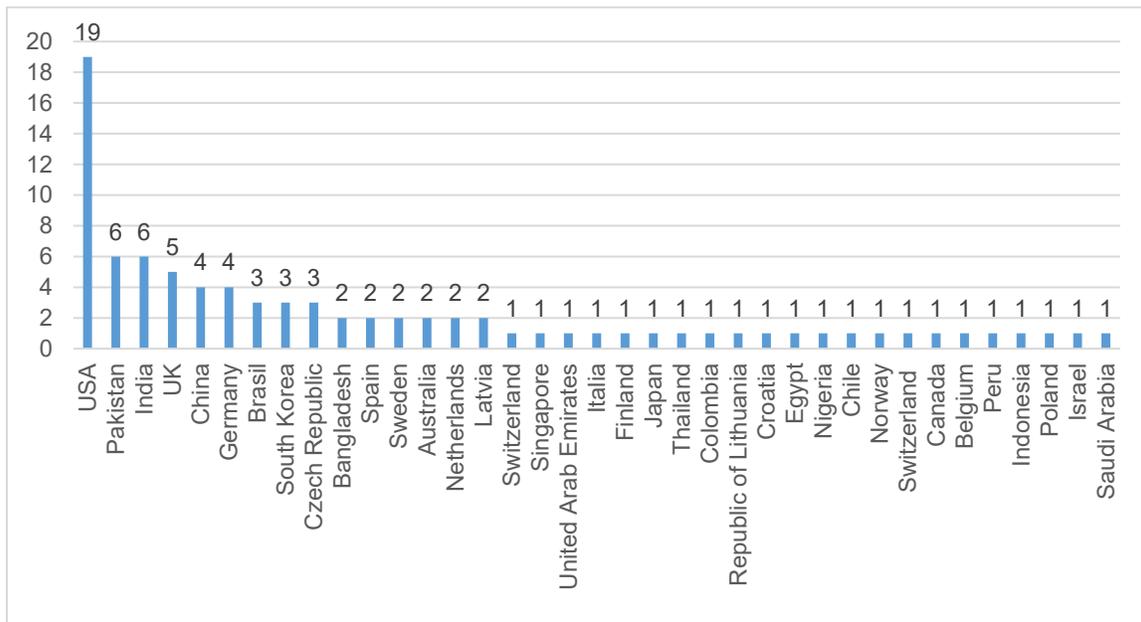

Fig. 3. Countries of the respondents

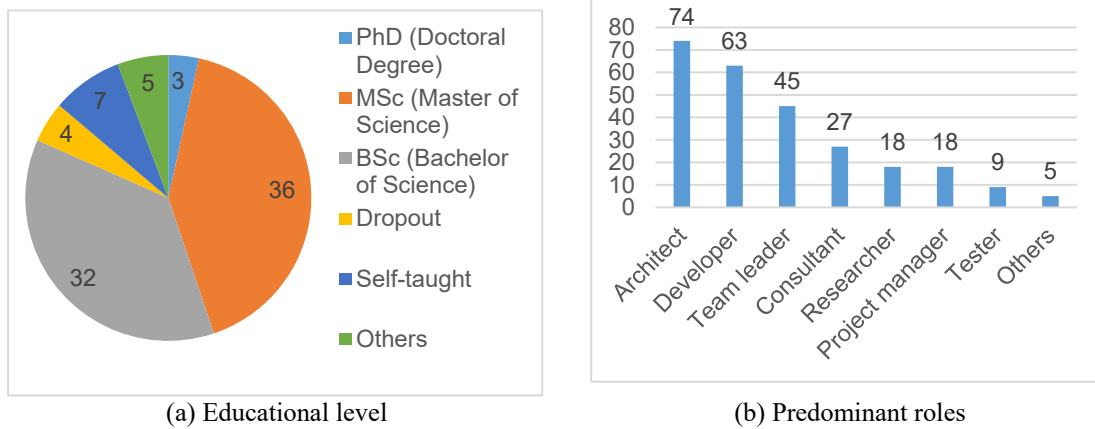

(a) Educational level                    (b) Predominant roles

Fig. 4. Educational level and predominant roles of the respondents

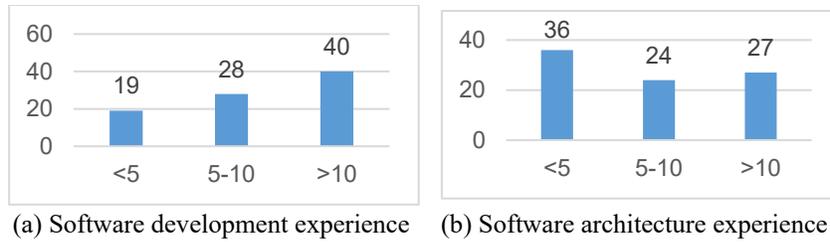

(a) Software development experience    (b) Software architecture experience

Fig. 5. Educational level and predominant roles of the respondents

As shown in Fig. 6(a), the respondents worked in different size of organizations. Only 12.6% (11 out of 87) worked in organizations with more than 250 employees. 66.7% (58 out of 87) respondents worked in organizations with 10-250 employees. Moreover, the respondents worked on diverse domains as shown in Fig. 6(b). The four most common domains were E-commerce, financial, telecommunication and healthcare. Most of the respondents worked in E-commerce (57.4%, 50 out of 87) and financial (57.4%, 50 out of 87), followed by Telecommunication (39.1%, 34 out of 87), Retail (33.3%, 29 out of 87), Healthcare (25.3%, 22 out of 87), Insurance (21.8%, 19 out of 87), and Embedded system (14.9%, 13 out of 87).

The demographic data demonstrates that the respondents are the representatives of the survey target population as they have experience in architecture design and software implementation as well as over a wide coverage of 37 countries, diverse work domains and roles and different size of organizations.

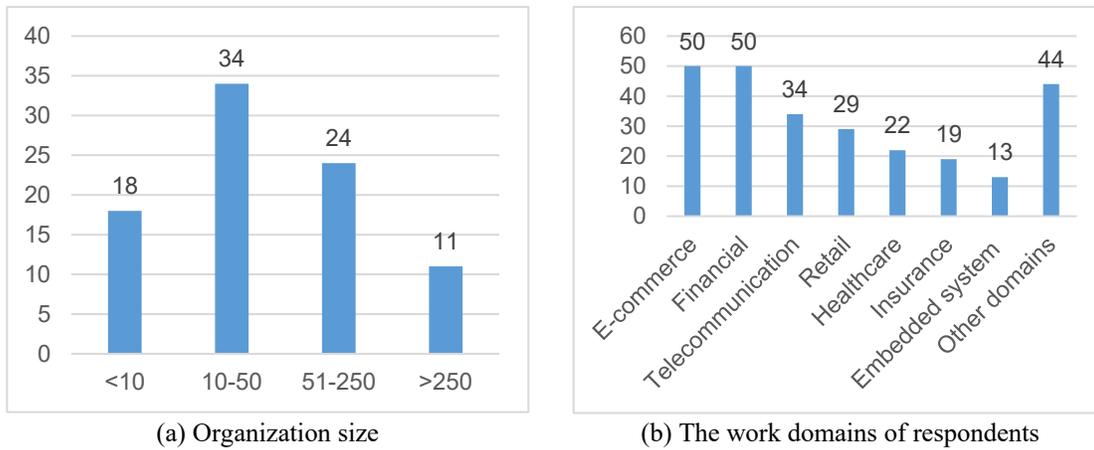

(a) Organization size    (b) The work domains of respondents

Fig. 6. The organization size and domains the respondents worked in

The demographic information in Table 4 shows that our interviewees cover five countries, six job roles, and diverse work domains with on average 86 months of software development experience and 62 months of architecture experience. Architect is the predominant role of the interviewees in their organizations. Moreover, most of the interviewees (6 out of 8) worked in an organization with more than 10 employees, which is consistent with the organization sizes distribution of the survey respondents.

Table 4. Demographical data of interviewees

| Interview Participant | Country | Main Role | Development Experience | Architecture Experience | Project Size | Work Domain |
|---|---|---|---|---|---|---|
| IP1 | China | Architect, Project manager, Team | 40 months | 37 months | <10 | Financial, Healthcare |

| | | leader, Developer | | | | |
|---|---|---|---|---|---|---|
| IP2 | China | Architect, Researcher | 31 months | 31 months | 10-50 | Telecommunication |
| IP3 | Australia | Architect, Developer, Consultant | 156 months | 84 months | 10-50 | Financial, Healthcare, Retail, Geographic Information System, Oil Gas & Petroleum, Learning Management |
| IP4 | Australia | Architect, Researcher | 60 months | 36 months | <10 | Healthcare, Cyber security |
| IP5 | Australia | Team leader, Architect, Researcher, Developer, Consultant | 168 months | 132 months | 50-250 | E-commerce, Financial |
| IP6 | Saudi Arabia | Researcher, Architect | 60 months | 24 months | 10-50 | Healthcare |
| IP7 | Italy | Team leader, Architect, Developer, Consultant, Researcher | 60 months | 100 months | 50-250 | E-commerce, Financial, Healthcare, Telecommunication |
| IP8 | Chile | Architect, Researcher | 110 months | 50 months | 10-50 | Financial, Healthcare |

## 5.2 RQ1: Understanding of the Relationships between SA and SC

We used five survey questions (Q9-Q13) to explore how practitioners understand the relationships between SA and SC. We initially asked them the question about whether there are certain features of relationships between SA and SC (Q9). All the respondents confirmed that the relationships between SA and SC do exist. Then the respondents were asked why they thought there are these relationships (Q10) and provide some examples (Q11). As described in Table 5, we identified five features from these relationships between SA and SC.

### 5.2.1 Transformability

Transformability means that SA can be transformed into SC. A majority of the respondents (69.0%, 60 out of 87) discussed this relationship which can be further classified into three subfeatures:

**Architecture leads or guides code**: 33 (37.9%, out of 87) respondents mentioned that architecture determines code structure or patterns, architecture guides development strategies, and architecture guides the implementation of code by defining architecture models (e.g., C4 model), architectural views, or architecture patterns (e.g., MVC patterns) with UML or architecture description languages. For example, two respondents stated that: "*The architecture defines the structure and organization of the code. Architecture defines the components, their structural organization, their interfaces, and the interactions required to perform the required overall system functions. It also defines the sequences of interactions required to execute overall system functions*"

**R27**. "*Architecture guides development strategies and design decisions. Architectures are changes that affect multiple stakeholders (other teams, governance, ...). Designs are changes that affect 1 team only*" **R1**.

**Code derived from the architecture**: 22 (25.3%, out of 87) respondents pointed out that code is generated from or implements the architecture specification, architecture diagrams, or reference architecture. As two respondents mentioned that: "*Cause any architecture needs implementation. Which is done by code*" **R25**. "*Patterns, practices and project-specific decisions instruct how architecture ought to be implemented in code*" **R55**.

**Architecture as an abstraction of code**: 5 (5.7%, out of 87) respondents reported that architecture provides the level of abstraction for code and architecture is the backbone of code. As two respondents observed that: "*Arch and code are on opposite sides of abstract/concrete spectrum*" **R7**. "*Architecture views reflecting the code aspect such as module views, code views, etc. directly represents the system's code structure*" **R13**.

### 5.2.2 Traceability

Traceability denotes that software architecture can be traced to source code of a system and vice versa. 26 (29.9%, out of 87) practitioners reported this relationship which can be further classified into three subfeatures:

**Dependency between architecture and code**: 12 (13.8%, out of 87) respondents indicated that architecture is dependent on code and vice versa, e.g., "*Coding depends on how domain model is created and how different systems interact with each other*" **R61**.

**Mapping between architecture and code**: 14 (16.1%, out of 87) respondents indicated that architecture should map to and communicate with code, and emphasizes that architectural elements (e.g., components) are corresponding to code elements (e.g., files) and vice versa. As one practitioner responded that: "*Mapping between architecture and code will make a system easy to explain. Impact of change is easily understandable*" **R72**.

Although several studies [25][26][62][63] proposed approaches or tools to support dynamic traceability between architectural elements and executing systems (e.g., runtime events), none of the survey respondents mentioned such dynamic traceability in their responses. The interview question IQ2.3 was designed in our interview to further investigate dynamic traceability from the perspective of practitioners. The results of IQ2.3 show that only three interviewees (**IP1**, **IP3**, **IP8**) considered dynamic traceability in practice, which can help them figure out the actual operation and execution of systems, such as how system services and processes interact at runtime. For other interviewees, they only considered static traceability between structure and source code, because their goals are to check whether the implementation realizes the functionalities of modules defined in architecture without considering how the modules interact and execute at runtime.

### 5.2.3 Consistency

Architecture consistency, considered by 24 (27.6%, out of 87) respondents, refers to the relationship between SA and SC that the intended architecture should be consistent with the implemented system, and this relationship can be classified into two subfeatures:

**Adherence of code to architecture**: 18 (20.7%, out of 87) respondents pointed out that the narrative code should adhere to a system's architecture. As two respondents explained "*Code adheres/diverges in various ways from architecture*" **R7**. "*Either code follows architecture or architecture follows code. In agile, the evolutionary architecture follows code. In a top-down, waterfall approach, code follows architecture*" **R26**.

**Architecture and code in sync**: This subfeature means that architecture and code should be changed synchronously due to e.g., new requirements. This subfeature is different from the subfeature "Dependency between architecture and code" in the feature Traceability, which emphasizes that the code implementation depends on the design of architecture or vice versa, while "Architecture and code in sync" is supported by "Dependency between architecture and code". 6 (6.9%, out of 87) respondents reported this relationship. As one respondent explained: "*My *personal goal is to keep design and implementation in sync throughout all the project's lifecycle*" **R56**.

### 5.2.4 Interplay

Interplay means that architecture quality influences code quality, and vice versa. 19 (21.8%, out of 87) respondents disclosed this relationship with two subfeatures:

**Architecture quality influences code quality**: 12 (13.8%, out of 87) respondents stated that the quality of architecture can impact the code quality and solid architecture leads to higher quality code. As one respondent observed that: "*Software architecture defines programming languages, layers, components, middleware etc., which heavily influence the relevant code patterns*" **R31**. This subfeature is different from the subfeature "Architecture leads or guides code" in the feature Transformability, because this subfeature refers to the situation that a poorly conceived and defined architecture can cause code flaws, defects, or anomalies, whereas the subfeature "Architecture leads or guides code" emphasizes that architecture decisions or patterns can determine the decisions of code structure and patterns.

**Code quality influences architecture quality**: Other 7 (8.0%, out of 87) respondents regarded that bad code decisions impact architecture and problems in code show drawbacks in architecture. As explained by one respondent "*There are plenty of them, but the more practical one at my sense is about Functional Relationship: Bad code impacts deeply high-level architecture imperatives*" **R43**.

### 5.2.5 Recovery

Architecture recovery is a process to extract architecture information from entities at the code level. 15 (17.2%, out of 87) respondents indicated that architecture can be recovered from code with the following subfeatures (i.e., reasons):

**Architecture embodied in code**: 8 (9.2%, out of 87) respondents stated that the elements in architecture design (e.g., tactics, components, and interfaces of the architecture) can be reflected in code. This subfeature is different from the subfeature "Code derived from the architecture" in the feature Transformability, in that the former highlights the recovery of architecture from code while the latter focuses on the generation of code from architecture, which are complementary to each other. For example, one respondent mentioned that "*architecture is the narrative that some code strives to embody*" **R7**.

**Identifying architecture from code**: 7 (6.9%, out of 87) respondents identified architecture from code artefacts, e.g., code structure or elements. This subfeature is different from the subfeature "Architecture embodied in code", because "Identifying architecture from code" highlights extracting the implemented architecture (e.g., components and their interactions) from code artefacts by using architecture recovery approaches and tools, while "Architecture embodied in code" focuses on the situation that architecture elements (e.g., components and tactics) can be reflected in code elements (e.g., interfaces and classes). For example, architectural elements can be generated by architectural recovery tools, as one respondent described "*you must be able to read a code, so for legacy system you start with code review, if there is a requirement document and architecture or similar documentation you should use it, other tools which can help are the tools which generates UML from source code*" **R11**.

Table 5. Features and subfeatures of the relationships between SA and SC known by the respondents

| Feature | Subfeature | Description | Example | Count (%) |
|---|---|---|---|---|
| Transformability | Architecture leads or guides code | Architecture determines code structure or patterns, and guides the implementation of code | *Architecture determines code by architecture diagrams.* (**R4**) *Architecture sets some guidance on what and how. Code is the actual how.* (**R1**) | 33 (37.9%) |
| | Code derived from the architecture | Code is generated from or implements the architecture specification, diagrams, or reference architecture | *With the support of UML, code can be produced from architectural diagrams and code changes can be folded back into the diagrams.* (**R30**) | 22 (25.3%) |
| | Architecture as an abstraction of code | Architecture provides the level of abstraction for code or the backbone of code | *Software architecture is an abstraction of the code and runtime structure.* (**R15**) | 5 (5.7%) |
| Traceability | Dependency between architecture and code | Architecture is dependent on code and vice versa | *Coding depends on how the domain model is created and how different systems interact with each other.* (**R61**) | 12 (13.8%) |
| | Mapping between architecture and code | Architectural elements (e.g., components) are corresponding to code elements (e.g., files) and vice versa | *The mapping between architecture and code will make a system easy to explain, the impact of change easily understandable, and less refactoring.* (**R72**) | 14 (16.1%) |
| Consistency | Adherence of code to architecture | Code should adhere to a system's architecture | *Because the code has to obey to the architecture, from architectural and design patterns to the simplest code, that's why there are tools and ways to determine the architecture adherence.* (**R19**) | 18 (20.7%) |
| | Architecture and code in sync | Architecture and code should be changed synchronously due to e.g., new requirements | *If your architecture is distributed, then the code must also be able to support the distributed system.* (**R77**) | 6 (6.9%) |
| Interplay | Architecture quality influences code quality | The quality of architecture can impact the code quality | *Either you architect and get good and predictable results with your code or you don't architect and you get an unpredictable pile of a mess of our code.* (**R12**) | 12 (13.8%) |
| | Code quality influences architecture quality | Bad code decisions impact architecture and problems in code show drawbacks in architecture | *Difficulties with code and problems in a system can show any drawbacks in architecture.* (**R60**) | 7 (8.0%) |
| Recovery | Architecture embodied in code | The elements in architecture design (e.g., tactics, components, and interfaces) can be reflected in code | *The code reflects software architecture components and structure definition.* (**R20**) | 8 (9.2%) |
| | Identifying architecture from code | Architecture can be identified from code artefacts (e.g., code structure or elements) | *Design patterns are one of them, the structure of the code (say the folder structure, and the file naming policies) will show them too (say, if you are using MVC pattern, you'll see the models, views and controllers in different folders, quite easy to find).* (**R8**) | 7 (6.9%) |

Once the respondents provided the relationships they knew between SA and SC to answer Q11, we further asked them whether they had ever considered the mentioned relationships (Q12 in Table 13) and to give examples of the relationships they frequently considered in their projects (Q13 in Table 13). From the answers to Q12 and Q13, we found that only five (5.7%, 5 out of 87) respondents neither considered the relationships they knew (answered in Q11), nor provided any other frequently considered relationships in their practical work. Combining with their demographic information, we discovered that these five respondents had at least two years' experience in SA and were working in the domain of Embedded system or Telecommunication. In contrast, as for 82 (94.3%, 82 out of 87) respondents who considered the mentioned relationships in their work, some of them (14.6%, 12 out of 82) had less than one years' experience in SA and were working in the domain of Financial system, Healthcare, or E-commerce. Furthermore, most of the 82 respondents (62.2%, 51 out of 82) considered Transformability, followed by Traceability (15.9%, 13 out of 82), Consistency (11.5%, 10 out of 82), Recovery (9.8%, 8 out of 82), and Interplay (6.1%, 5 out of 82). Note that, the respondents only considered one of their known relationships (Q11) to use in practice when answering Q13, therefore the number of the relationships that are known (in Table 5) by the respondents and that are considered by them in practice are different.

These five relationships were further investigated through our interview (IQ2.1.1-IQ2.1.6) to analyze whether and how practitioners use each relationship. Note that the relationship Interplay has a difference when considering the influence direction, we regarded this relationship as two relationships: Interplay (architecture to code) and Interplay (code to architecture). Table 6 shows that all the interviewees used Transformability, Traceability, Consistency, and Interplay (architecture to code) in their projects. Transformability (5 out of 8) and Consistency (5 out of 8) were frequently used by most interviewees. Moreover, only three interviewees (**IP1**, **IP6**, **IP8**) considered that poor code quality can influence architecture quality. For example, **IP1** mentioned that: "*Architecture is designed to be of high performance and availability. If code is not implemented appropriately with some high availability functionalities, the architecture is impacted, because code cannot implement what the architecture is designed for*". Except **IP2** and **IP4**, the other interviewees used Recovery relationship in their projects. For example, **IP3** discussed that "*We used IntelliJ normally, because I used to do Java coding in my experience. So IntelliJ opens the source files and can generate architecture diagrams which navigate to different classes*".

Table 6. Using of each relationship feature by the interviewees

| Interviewee<br>Relationship | IP1 | IP2 | IP3 | IP4 | IP5 | IP6 | IP7 | IP8 |
|---|---|---|---|---|---|---|---|---|
| **Transformability** | √ | √* | √* | √* | √* | √* | √ | √ |
| **Traceability** | √ | √ | √ | √ | √ | √ | √ | √* |
| **Consistency** | √* | √ | √* | √ | √* | √ | √* | √* |
| **Interplay (architecture to code)** | √ | √ | √ | √ | √ | √ | √ | √ |
| **Interplay (code to architecture)** | √ | × | × | × | × | √ | × | √ |
| **Recovery** | √* | × | √ | × | √* | √ | √* | √* |

Note: '√', '×', and '*' denote that certain feature of relationship is used, not used, and frequently used by interviewees, respectively.

### 5.2.6   Interrelationships between the SA and SC relationships

The interviewees were also asked in the interview question IQ2.1.8 about the interrelationships between the SA and SC relationships. Table 7 illustrates that Traceability and Recovery relationships can support each other; Transformability can support two relationships (i.e., Traceability and Consistency). For example, one interviewee mentioned that: "*The idea of architecture is trying to make traceability with source code, because of the recovered architecture, you will see the big picture of architecture. But for us, it's also important to try to show in somehow how that recovered architecture can show the traceability with the source code*" **IP1**. We also found that no interviewees considered that Transformability and Interplay relationships can be supported by other relationships.

Table 7. The interrelationships between the SA and SC relationships

|  | Transformability | Traceability | Consistency | Interplay | Recovery |
|---|---|---|---|---|---|
| Transformability |  | IP2, IP3, IP7 (→) | IP1, IP2, IP6, IP7 (→) |  |  |
| Traceability |  |  | IP4, IP5 (→) |  | IP4 (→) |
| Consistency |  |  |  |  |  |
| Interplay |  |  |  |  |  |
| Recovery |  | IP8 (→) | IP1 (→) |  |  |

Note: '→' denotes that the relationship in the row can support the relationship in the column

### 5.2.7 Architectural concepts and knowledge to support the SA and SC relationships

The interview questions IQ2.2.1 and IQ2.2.2 were used to investigate the architectural concepts and knowledge which are considered important by the interviewees to support the SA and SC relationships. The interviewees considered architecture views (7 out of 8), architecture decisions (3 out of 8), architecture patterns (2 out of 8), and layers (1 out of 8) as important architectural knowledge to support the relationships between SA and SC. For example, one interviewee answered that: "*Fine-grained architectural models with architectural views can be easily linked to the source code. If the architecture can be represented with different views (e.g., deployment views, structure views, behavior views), that will certainly help us build a good relationship between architecture and code*" **IP4**. The responses of IQ2.2.2 also reveal that most of the mentioned architectural knowledge is used with UML diagrams by the interviewees to support the SA and SC relationships, as one interviewee mentioned: "*Deployment diagram in UML is very important. We should first draw the overall deployment diagram and the whole chain of network traffic inbound and outbound to check whether the deployment diagram can support high concurrent access*" **IP1**. Two interviewees (**IP2**, **IP6**) also mentioned that architectural knowledge can be used to support automatically transforming models into code. For example, **IP2** mentioned that "*Architecture views are mainly reflected in the early design of the architecture model to verify that architecture views are designed correctly. We then automatically convert architecture views into the test code*".

Moreover, most interviewees (7 out of 8) responded to the interview question IQ2.2.3 that architecture decisions, as an important type of architectural knowledge, impact implementation decisions, because architecture decisions can determine code structure and code changes. For example, one interviewee answered that "*When you have, for example, implemented architecture decisions, every decision that you made in their architecture, have to be implemented in the source code. So, for example, you decide to implement a cryptography in the communication between two components. Of course, you have to see the cryptography in the source code*" **IP8**. Only one interviewee from the seven interviewees above (**IP1**) argued that wrong implementation decisions compromise architecture decisions. For example, one interviewee mentioned that "*For some systems with high performance requirements, it is supposed to add a layer of caching in the architecture for high performance. If not, it will result in slow throughput rate. Architecture decision has been designed with high-performance processing. Therefore, if the code is designed as non-high performance processing, the performance of the architecture will be largely impacted and the throughput rate cannot be improved*" **IP1**.

**Key Findings**

***Finding 1.*** *Five relationships between SA and SC are known by practitioners. Among these relationships, Transformability, Traceability, and Consistency are the most frequently mentioned and considered.*

***Finding 2.*** *Transformability, Traceability, and Consistency are frequently considered and used relationships between SA and SC by practitioners.*

***Finding 3.*** *There are certain interrelationships between Transformability, Traceability, and Consistency, but Transformability and Interplay cannot be supported by other relationships.*

***Finding 4.*** *Architecture views, architecture decisions, architecture patterns, and layers were considered as important architectural knowledge to support the SA and SC relationships.*

### 5.3 RQ2: Approaches and Tools for Identifying, Analyzing, and Using the Relationships between SA and SC

The respondents were asked to rank the importance of identifying, analyzing, and using the relationships between SA and SC in software development (Q14). This ranking illustrates the respondents' perception of how valuable it is for identifying, analyzing, and using the relationships between SA and SC. As shown in Fig. 7, a majority of the respondents (89.7%, 78 out of 87) considered it as important or very important to identify, analyze, and use the relationships between SA and SC, with 65.6% (57 out of 87) respondents considering it as very important. But still, two (2.3%, out of 87) respondents have no idea of the importance and one (1.1%, out of 87) respondent stated that it is unimportant to focus on the relationships between SA and SC. It should be noted that although five (5.7%, out of 87) respondents did not consider the relationships in their projects as mentioned above, they considered it important to identify, analyze, and use the relationships between SA and SC.

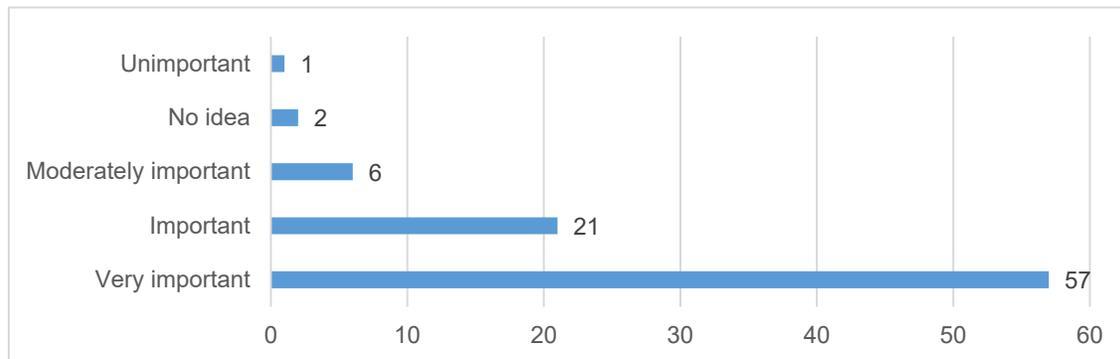

Fig. 7. Importance of identifying, analyzing, and using the relationships between SA and SC

#### 5.3.1 Approaches for identifying, analyzing, and using the relationships between SA and SC

We asked the respondents and interviewees how they identified, analyzed, and used the relationships between SA and SC (Q15, Q16, IQ2.1.1-IQ2.1.6, IQ3.1, IQ3.2). As listed in Table 8, we classified the approaches into three types: top-down integration, bottom-up integration, and other specific approaches. For top-down integration, it means how to identify, analyze, and use the relationships between SA and SC from the high-level SA. Most practitioners (71.6%, 68 out of 95 (87+8)) mentioned the top-down integration approach which was mainly used in the waterfall development process with which SA is determined up-front before SC development begins. Among them, 45.3% (43 out of 95) practitioners adopted Pattern-Oriented Design (POD) to structure and develop the systems. The specific patterns utilized by the respondents are C4 model, MVC, Event Sourcing, and so on. Moreover, 21.1% (20 out of 95) and 15.8% (15 out of 95) practitioners, respectively employed Service-Oriented Development (SOD) and Domain-Driven Design (DDD) to define service-oriented architecture (e.g., microservice architecture) and domain models, and implement these designs at the code level. As one respondent described that: "*Coding depends on how the domain model is created and how different systems interact with each other*" **R61**. By using POD, SOD, and DDD for maintaining the relationships between SA and SC, the respondents firstly understood the Architecturally Significant Requirements (ASRs) and employed architecture patterns for addressing the ASRs. Once architecture patterns were determined, they implemented these patterns consistently in code. As one respondent mentioned: "*Because I'm primarily working in microservices architecture, I mostly consider the types of code patterns that connect distributed systems (i.e. REST API design)*" **R47**. For automatic code generation, Model-Driven Development (MDD) was only adopted by 4 practitioners (4.2%, out of 95) to automatically transform architecture models to code, which can also build the Traceability links and ensure Consistency between SA and SC.

Furthermore, 30 (31.6%, out of 95) practitioners adopted bottom-up integration to maintain the relationships between SA and SC from the code level, which usually happens in agile methods and iterative and incremental software development. Among them, 14 (14.7%, out of 95) practitioners employed Scrum or pair programming to build architecture interactively and incrementally and evolve the architecture in parallel with code. As one respondent mentioned: "*I do not use any formal method. My favourite approach is agile: start with code and adapt and refactor it into a well-structured software architecture*" **R26**. For the other 13 (13.7%, out of 95) practitioners, they managed the relationships between SA and SC by code review with team members or automated code review tools to debug code in order to establish the relationships between SA and SC. Moreover, reverse-engineering approaches were used by six practitioners (6.3%, out of 95) to extract UML diagrams from code.

Lastly, 37 (38.9%, out of 95) practitioners adopted specific approaches to support the relationships between SA and SC which can be employed in both top-down and bottom-up integration. 25 (26.3%, out of 95) practitioners used architecture modelling or visualization with UML diagrams to document and check architecture documents in conjunction with code. In addition, 15 (15.8%, out of 95) practitioners conducted peer review discussion or used informal verbal and written communication to explain and analyze the architecture to be implemented at the code level. Six practitioners (6.3%, out of 95) manually documented trace links between SA and SC by using architectural and code element names, code comments, and traceability tables and files. Another five practitioners (5.3%, out of 95) ensured the consistency between SA and SC by manually changing architecture and code iteratively. On top of that, two practitioners discussed how to use ATAM [66], TOGAF [67], and Spring framework for conceptualizing and analyzing the relationships between SA and SC. Two practitioners also used logs to trace call chains in code to actual execution of architectural behavior (i.e., dynamic traceability). Only one practitioner reported that Design Structure Matrix (DSM) was employed to recover the dependency structure of architecture and trace code elements to architecture dependency.

Table 8. The approaches for identifying, analyzing, and using the relationships between SA and SC

| Type | Subtype | Example | Count (%) |
|---|---|---|---|
| Top-down integration | Pattern-Oriented Design (POD) | *I choose design patterns, practices and project-specific decisions to instruct how architecture ought to be implemented in code.* (**R55**) | 43 (45.3%) |
| | Domain-Driven Design (DDD) | *I analyzed the load and domain of the application to decide the type of relationship to maintain.* (**R77**) | 20 (21.1%) |
| | Service-Oriented Development (SOD) | *Queue system leads to reactive programming. Microservices lead to orchestration code.* (**R23**) | 15 (15.8%) |
| | Model-Driven Development (MDD) | *In one of our projects, we used a model-driven approach which includes four stages: create diagrams or models, automatically convert to source* | 4 (4.2%) |
| | | *code, manually change and check the architecture documents, then accordingly change the code iteratively.* (**IP4**) | |
| Bottom-up integration | Agile development | *Working in parallel (agile/scrum), a code written by one team might not be good for another team or can be conflicting. This clearly needs a design relation between architecture and code.* (**R2**) | 14 (14.7%) |
| | Code review | *The first approach is code review, then analysis of the test cases, identification of the hot spots in architecture design or places where code should be improved.* (**R11**) | 13 (13.7%) |
| | Reverse-engineering UML diagrams from code | *The logical diagrams that a class starts from data usually are generated if we need. That is generated by reverse engineering code.* (**IP5**) | 6 (6.3%) |
| Specific approaches | Architecture modelling and visualization | *Using architecture components models based on requirements and checking its viability with PoCs and assuring its alignment using software frameworks, code templates and standards.* (**R10**) | 25 (26.3%) |
| | Peer review discussion | *You have to study implementation and specification again, ask a few experts, and if all agree, then you make the claim that the architecture is correct and that you provided an implementation of that architectural specification.* (**R15**) | 15 (15.8%) |
| | Documenting traces between architecture and code | *The names of the architectural modules correspond to the code, such as package names, class names, and method names; Calling chains in architecture diagrams are traced to chains of interfaces and method calls.* (**IP1**) | 6 (6.3%) |
| | Manually changing architecture documents and code iteratively | *Most of time, we manually change and check the architecture documents, then accordingly change the code iteratively.* (**IP4**) | 5 (5.3%) |
| | ATAM | *Finally, I use the ATAM (Architecture Tradeoff Analysis Method) technique to analyze the architecture decisions and their possible negative consequences in the materialization of the software that is the code.* (**R59**) | 2 (2.1%) |
| | TOGAF | *Using the TOGAF framework which again is a framework which is mainly used to follow the approach of enterprise architecture development during which there are certain phases of ADM (Architecture Development Method) to conceptualize the relationships between software architecture and code.* (**R70**) | 2 (2.1%) |
| | Spring framework | *For example, I develop a web application through a framework such as Spring Framework, spring boots explain the ecosystem. So in that case, you use the framework, and the framework uses the, for example, model view controller architecture from the side, so the transformation from software architecture to source code is done manually.* (**IP7**) | 2 (2.1%) |
| | Using logs to build dynamic traceability | *For our current project, the practical execution is trace about call chains. Logs are used to trace parameters of call chains and analysis of operational aspects (e.g., traffic in and out) to the* | 2 (2.1%) |
| | | *analysis of high throughput rate of the architectural structures.* (**IP1**) | |
| | Recovering and building the traces by Design Structure Matrix (DSM) | *DSM is important for traceability, because, for example, you need to see the relationships between packages or source code or anything that you can represent in the DSM.* (**IP8**) | 1 (1.1%) |

**Key Findings**
***Finding 3.*** *The type of approaches adopted by most respondents for identifying, analyzing, and using the relationships between SA and SC are the top-down integration approaches, such as Pattern-Oriented Design (POD), Domain-Driven Design (DDD), and Service-Oriented Development (SOD).*
***Finding 4.*** *Dedicated approaches proposed by researchers (as discussed in Section 3.1) are less adopted by practitioners to identify, analyze, and use the relationships between SA and SC.*
***Finding 5.*** *Dedicated approaches (e.g., MDD and DSM) from the literature are used by practitioners to automatically support certain relationships (e.g., Traceability, Consistency).*

### 5.3.2 Tools for identifying, analyzing, and using the relationships between SA and SC

We also asked respondents and interviewees about the tools they used to identify, analyze, and use the relationships between SA and SC (Q15, Q16, IQ3.2). As described in Table 9, 34 (35.8 %, out of 95) practitioners reported that the most frequently used tools in their daily work are knowledge and experience. As one respondent stated: "*The most reliable one that you should use, is your experiences. When you found struts-config.xml in the WEB-INF folder, then it must be a Java EE War project, and it is based on the MVC design pattern and it will use struts as the framework, and you'll see many java files naming end with Action.java*" **R18**. Next, 26 (27.4%, out of 95) practitioners used architecture visualization and modelling tools to discuss and manually record the relationships between SA and SC. Furthermore, 15 (15.8%, out of 95) respondents only analyzed the relationships between SA and SC in integrated development tools, e.g., Visual Studio, because these tools can integrate architecture diagrams with developed code and validate code with the architecture diagrams, as one respondent mentioned "*To make sure that code doesn't conflict with its design, validate your code with dependency diagrams in Visual Studio*" **R81**. 10 (10.5%, out of 95) practitioners employed static code analysis tools to extract the dependency structure of code for representing SA during code inspection and review. Last, 7 (7.4%, out of 95) practitioners used dedicated architecture recovery tools (e.g., Codecrumbs, Ctags, Understand) to generate dependency graph or UML diagrams from source code.

Table 9. The tools for identifying, analyzing, and using the relationships between SA and SC

| Type | Subtype | Example | Count (%) |
|---|---|---|---|
| Knowledge and experience | Knowledge in books, Intuition, Team members skills | *No tools needed; this is just what I have seen in my 38 years of being involved in software engineering and our mind.* (**R10**) | 34 (35.8%) |
| Architecture visualization and modelling tools | MS Visio, Draw.io, C4 Model, Whiteboard, Enterprise Architect, MS PowerPoint, ObjectAid, LucidCharts, OmniGraffle, Dia Diagram Editor | *Usually, I make a class diagram first before code it. the tools that I am used are DIA diagram and draw.io after creating a class diagram then I implement it on a real project. The tools are depending on kinds of project. For example, I use android studio to implement the architecture on android project.* (**R48**) | 26 (27.4%) |

| Integrated development tools | IntelliJ IDEA, Visual Studio, Rational Software Architect, Structure101 Studio, Lattix, Rational Suite | *IntelliJ plugins allow you to see the relationships between the classes and based on this to build the overview of architecture.* (**R54**) | 15 (15.8%) |
|---|---|---|---|
| Static code analysis tools | FindBugs, Sonatype, JIRA, TFS, Code quality management tools (e.g., Eslint, Prettier) | *I use low-level code quality tools like Eslint, Prettier, only drawing architecture as a UML diagram, usually in Draw.io. When I came into a new existing project, I will take a look into docs and repositories and make an overall representation in my head.* (**R63**) | 10 (10.5%) |
| Architecture recovery tools | CodeCrumbs, Ctags, Understand, Diagram generation tools | *Other tools which can help are the tools which generate UML diagrams from source code.* (**R11**) | 7 (7.4%) |

**Key Findings**
***Finding 6.*** *Practitioners mostly utilized their knowledge and experience to identify, analyze, and use the relationships between SA and SC, and dedicated tools proposed by researchers (as discussed in Section 3.1) are less used.*

## 5.4 RQ3: Benefits and Limitations of Identifying, Analyzing, and Using the Relationships between SA and SC

### 5.4.1 Benefits of identifying, analyzing, and using the relationships between SA and SC

We classified the benefits into several Quality Attributes (QAs) that can be improved by identifying, analyzing, and using the relationships. We combined the taxonomy of QAs defined in the ISO 25010 standard [53] and other QAs mentioned by Bass *et al*. [6]. As shown in Fig. 8, More than half of the respondents (56.3%, 49 out of 87) claimed that the system maintainability can by improve by identifying, analyzing, and using the relationships between SA and SC because it facilitates the circularization or modularization of code. Also, many respondents (24.1%, 21 out of 87) considered that reliability can be improved by minimizing the code bugs and validation of architecture by identifying, analyzing, and using the relationships. Some respondents mentioned that performance can be increased with less and clean code. Moreover, the respondents also stated that we can better understand architectures, reduce the complexity of implementation and reduce the overall cost to deliver a system so that the complexity, understandability and productivity of a system get improved.

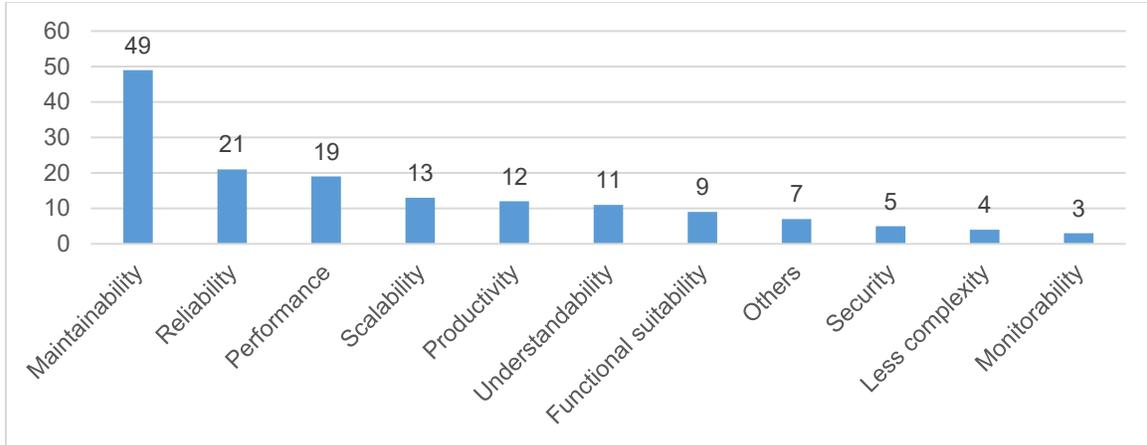

Fig. 8. The benefits of identifying, analyzing, and using the relationships between SA and SC

The interview question IQ4.1 was formulated to investigate how quality attributes of systems can be improved through identifying, analyzing, and using the relationships between SA and SC. As shown in Table 10, all interviewees claimed that Maintainability can be improved with the support of Consistency, Traceability, and Recovery relationships. As **IP4** mentioned that: "*If you want to have a higher maintainability, the good links can make you easily add somethings in architecture and accordingly visualize and locate the changes in code*". Two interviewees reported that Reliability can be improved with the support of Transformability and Consistency relationships. As **IP1** explained that: "*Similarly, if the architecture is consistent with the code, the maintainability and reliability code can be guaranteed through checking the architecture diagram*". Moreover, one interviewee (**IP3**) argued that Reusability and Scalability can also be improved with the support of Consistency relationship, as he stated that: "*A good architecture also enables us to write code in consistent with architecture that is reusable. We can write a reusable component. With a good architecture, we can deploy those components on a scalable environment*" **IP3**.

Table 10. The quality attributes improved by the relationship between SA and SC

| Relationship<br>Quality attribute | Transformability | Traceability | Consistency | Interplay | Recovery |
|---|---|---|---|---|---|
| **Maintainability** | IP2 | IP3, IP4, IP5, IP6, IP7, IP8 | IP1, IP3 | | |
| **Reliability** | IP2 | | IP1 | | |
| **Reusability** | | | IP3 | | |
| **Scalability** | | | IP3 | | |

**Key Findings**
*Finding 7. Most practitioners realized that quality attributes (especially maintainability and reliability) of systems are improved through identifying, analyzing, and using the relationships between SA and SC.*

### 5.4.2 Limitations of identifying, analyzing, and using the relationships between SA and SC

We collected and classified the difficulties in identifying, analyzing, and using the relationships between SA and SC as presented in Table 11. Most of the developers are concerned about the budget and cost involved in detecting and refactoring architectures. Before considering the relationships, they made a trade-off between effort and benefit because it uses manpower and time, especially in a tight software development project schedule. 13 (14.9%, out of 87) respondents had trouble in analyzing the relationships between SA and SC because they lack the available architecture documentation. Other 13 (14.9%, out of 87) respondents explained that the limitation is that enough knowledge and experience is needed to design architecture well and look for these relationships. Moreover, the gap between architecture and code can hinder the analysis of these relationships, in which case architects should communicate with developers and understand what types of programming languages the developers used for this architecture of the project. Besides that, some respondents worried about the lack of tools and team skills limitation on well implementing the architecture. There are still a few respondents that are unaware of any limitations of identifying, analyzing, and using the relationships between SA and SC.

Table 11. The limitations of identifying, analyzing, and using the relationships between SA and SC

| Difficulty | Example | Count (%) |
|---|---|---|
| Cost and effort | *There are challenges mainly in complex software projects due to big codebase, and also regard a tight software development project schedule.* (**R17**) | 35 (40.2%) |
| Lack of architecture documentation | *Professional software must have proper architecture, unfortunately, this not always the case in practice.* (**R11**) | 13 (14.9%) |
| Inexperience/Lack of knowledge | *Architecture happens in the beginning when the least knowledge is available. This is bad and should be avoided.* (**R1**) | 13 (14.9%) |
| The gap between architecture and code | *Concepts in architecture can often NOT be identified automatically in code.* (**R49**) | 11 (12.6%) |
| Lack of tools | *There is insufficient tooling available. No tool can tell whether the code actually implements a specification.* (**R15**) | 9 (10.3%) |
| Team limitations | *Limitations are the right strength and right capability of the architect team.* (**R76**) | 8 (9.2%) |
| Unawareness | *I think there is no limitation. I have to implement good architecture into any software development project.* (**R48**) | 5 (5.7%) |

**Key Findings**

*Finding 8. The cost and effort are the main obstacle that prevents practitioners from identifying, analyzing, and using the relationships between SA and SC, and lack of architecture documentation and knowledge is another two significant inhibiting factors.*

### 5.4.3 Cost-benefit analysis of identifying, analyzing, and using the relationships between SA and SC

Because cost and effort are the main difficulties in identifying, analyzing, and using the relationships between SA and SC, the interview question IQ4.2 was formulated to investigate how practitioners analyze the cost and benefit of identifying, analyzing, and using the relationships between SA and SC. Most of the interviewees (6 out of 8) considered that the benefits outweigh the costs according to their experience in development, as one interviewee explained that: "*Analyzing the relationship between code and architecture is an essential process in software design and development. In my project experience, benefits certainly outweigh costs, but it is difficult to accurately analyze the costs and benefits*" **IP1**. Only one interviewee (**IP5**) considered that the costs outweigh the benefits according to his experience and he mentioned that "*So the obviously, there are benefits, but for example, in my opinion, the cost would be way too much, as compared to benefits. No organization would like to invest that much effort, because that might have some return after five years*". Furthermore, one interviewee (**IP7**) did not consider and analyze the costs and benefits and he explained that "*The cost and effort was very simple and I didn't use an approach or method for defining constant benefits, because the systems were very simple*".

# 6 Discussion

In this section, we summarize and interpret the main findings of our study. We discuss the implications of the findings for practitioners and researchers.

According to the survey results, we observe that the relationships between SA and SC are considered important and have been taken into consideration by practitioners in the software development lifecycle. There is also a general perception that most practitioners usually employ several approaches such as POD, DDD and code review, with the support of architecture modelling tools and experience to identify, analyze and use the relationships between SA and SC, which indicates that few dedicated approaches and tools proposed by researchers for different relationships between SA and SC have been adopted by practitioners, in spite of the known benefits, i.e., the maintainability and reliability of systems can be improved by identifying, analyzing and using the relationships between SA and SC. Meanwhile, there are difficulties in identifying, analyzing and using the relationships between SA and SC.

## 6.1 Summary of the Main Findings

### 6.1.1 Perception of practitioners on the relationships between SA and SC

The survey results of Q12 and Q14 indicate that nearly all the respondents (more than 90%) agreed upon the critical importance of identifying, analyzing and using the relationships between SA and SC in software development. This is not a surprising result considering that all the respondents were conscious of the potential benefits of the knowledge of the SA and SC relationships for a system as mentioned in Section 5.4.1. We also observe from the results of Q11 and Q13 that most of the respondents considered and used relationships between SA and SC are Transformability, Traceability and Consistency. When it comes to the relationships between SA and SC, more than half of the respondents (69.0%, 60 out of 87) concerned about Transformability that architecture can be implemented in code. One third of the respondents (29.9%, 26 out of 87) took Traceability into consideration that architecture depends on code and vice versa. A similar number of the respondents (27.6%, 24 out of 87) considered Consistency that SC should adhere to SA or SA and SC should be kept in sync. Only a small number of the respondents mentioned other two types of relationships, i.e., Interplay (21.8%, 29 out of 87) and Recovery (18.4%, 16 out of 87). These results indicate that most practitioners considered the explicit relationship (Transformability) that code is generated from architecture, which is reasonable according to the results of approaches (see Section 5.3.1) that most practitioners used Top-down integration approaches (i.e., from architecture to code). In contrast, fewer respondents thought about the relationships from the perspective of code that architecture can be recovered from the code and the quality of architecture can be impacted and evaluated by the quality of code.

Except for three interviewees (3.1%, 3 out of 95), most practitioners only mentioned static traceability between structural architecture elements and code elements without further considering dynamic traceability between the behavior of SA and SC [20], because their main goal is to ensure that the functionalities of architecture components can be implemented in code. In addition, 22 respondents realized that the quality of architecture can be impacted and evaluated by analyzing the quality of code, but they did not specifically describe which kinds of code problems (e.g., code smells or vulnerabilities) can be the indicators of architecture problems (e.g., architectural smells) [43][45].

Certain interrelationships between the SA and SC relationships are mentioned in the literature; when discussing these relationships, e.g., architecture consistency, researchers usually consider to use, e.g., code generation from architecture, architecture recovery, or architecture traceability as a technique to check architecture consistency and address architectural erosion [21][23]. The above interrelationships were also observed in our interviews, because the interviewees highlighted that Transformability, Traceability, and Recovery relationships can support Consistency relationship between SA and SC. Moreover, several studies [42][43][44] observed that Traceability between architecture and code can facilitate analyzing the impact of code smells on architecture smells (Interplay relationship), but no interviewees reported this interrelationship. This observation is explainable by the result that Interplay relationship is not frequently used by the practitioners in their projects.

### 6.1.2 Approaches and tools used for identifying, analyzing, and using the relationships between SA and SC

The survey results about approaches (see Section 5.3.1) and tools (see Section 5.3.2) indicate that most of the respondents used the top-down development approaches, such as pattern-oriented software architecture and waterfall development process, to identify, analyze, and use the relationships between SA and SC. For these respondents, they considered that architecture comes first, then the code implements and follows the architecture. Therefore, they mostly utilized architecture documentation and modelling as well as corresponding tools (e.g., MS Visio, Draw.io, C4 model or Whiteboard) to identify, analyze and use the relationships. But part of the respondents considered the relationships at the code level, as they used code review [57] and static code analysis tools (e.g., Sonatype, Findbugs) to recover and validate the implemented architecture. Passos *et al.* [27] also reported that architecture conformance can be checked with the support of static analysis tools (e.g., Klockwork, JDepend), but none of these tools were mentioned by the respondents in our survey. Moreover, some respondents chose the professional knowledge and experience as reliable approaches and tools to identify, analyze and use the relationships between SA and SC.

The results of approaches and tools also indicate that most of the practitioners who participated in our study employed ADLs (e.g., UML diagrams), domain-specific languages in DDD, and informal architectural models (e.g., boxes and lines) to support code generation for narrowing the gap between SA and SC. These traditional ways to communicate SA with SC are done manually according to practitioners' experience and understanding of a system's architecture. Unfortunately, as mentioned by Woods and Rozanski [41], it cannot really address the lack of architectural information in the implementation and the danger of architecture drift by using these methods for code generation or building traceability between SA and SC manually. Moreover, as pointed out by Passos *et al.* [27], although ADLs are used for checking architecture conformance, the implementation of ADLs is restricted to the systems that use mainstream programming languages.

Three architecture analysis tools (i.e., Structure101, Bahaus, Sotograph, and Lattix) can be used for checking architecture conformance [27], but the practitioners in our study only used Structure101, and Lattix for checking the consistency between SA and SC. In the interviews by Ali *et al.* [21], Structure101 was also adopted by one out of 19 participants to address architecture consistency. Whilst employing architecture evaluation techniques (e.g., ATAM) or tools (e.g., Structure 101) can build certain relationships between SA and SC, the practitioners still failed to address the gap between SA and SC since these techniques cannot integrate architecture specifications seamlessly into implementation code for unifying SA with SC [40]. Only one interviewee used Design Structure Matrix (DSM) supported by Lattix which is one of the three static architecture conformance checking techniques [27]. But as mentioned by Passos *et al.* [27], DSM with Lattix is insufficient to express even simple architecture constraints, because its design rules language is rather limited. More importantly, it cannot support dynamic traceability and consistency between SA and SC. For the approaches and tools supporting dynamic traceability, only two practitioners in our study (2.1%, 2 out of 95) used logs to support dynamic traceability between SA and SC, and no specific approaches or tools for dynamic traceability (e.g., PANDArch [62]) presented in literature (see Section 3.1) were adopted by the practitioners. One potential reason is that it is challenging to capture relevant and useful data (e.g., method invocations and dynamic linking) from system execution for tracing and validating architecture constraints while keeping a minimized impact to system performance [62].

These results reveal that a few practitioners applied dedicated architecture recovery, traceability, and consistency approaches and tools proposed in the literature, such as those mentioned in Section 3, which is partly in line with the observation by Ali *et al.* [21] that 19 participants, except one who used a formal architecture consistency tool proposed in the literature, employed informal practices (e.g., communication channels between architects and developers) to maintain architecture consistency.

### 6.1.3 Benefits and difficulties of identifying, analyzing, and using the relationships between SA and SC

As shown in Fig. 7 and Fig. 8, most of the respondents (78 out of 87, 89.7%) still believe that understanding and use of the relationships between SA and SC is important for improving a system's qualities with the emphasis on maintainability, reliability and performance. But in their work, some factors may impede them from identifying, analyzing and using the relationships between SA and SC. As shown in Table 11, the main barrier they perceive to identify, analyze and use the relationships in practice is the cost and effort needed for bridging the gap between SA and SC. The problem, also reported by Keim and Koziolek [51], that architecture documentation does not exist, heightens participants' reluctance to adopt architecture consistency approaches and tools. It costs a lot of time and effort to recover the architecture from code and build up the relationships between them. Moreover, a lack of tools is another impediment that prevents practitioners from considering the relationships between SA and SC. In addition, several respondents show their unawareness of the relationships between SA and SC, which is probably explained by the fact that identifying, analyzing, and using the relationships are not treated as a main activity especially in agile development [52].

Moreover, architecture inconsistency for short-term gains is a typical symptom of technical debt [64], which will be paid with more effort in the future. Therefore, the ignorance of the relationships between SA and SC could bring short-term gains for fast development and delivery, but more effort is required to maintain systems and repay this technical debt in the long run [65]. All the interviewees realized the benefit that the Maintainability can be improved by analyzing and using the relationships between SA and SC. Although most interviewees considered that the benefits outweigh the costs according to their experience, currently no effective approaches have been available to quantify the benefits and costs. Therefore, the unclear cost-benefit quantification and time pressure for fast delivery hinder the consideration of identifying, analyzing, and using the relationships between SA and SC, which could accumulate architectural technical debt unconsciously [65].

## 6.2 Implications for Researchers and Practitioners

**For practitioners** - they shall consider the five relationship features between SA and SC as fundamental elements when they design an architecture and implement it accordingly. Practitioners also need to pay more attention to two relationships between SA and SC, i.e., Interplay and Recovery. These two relationships are easier to be overlooked (less discussed than other relationships according to the survey results) in practice with the pressure of time and budget. However, Interplay, as mentioned in Section 3, is essentially important to evaluate architecture quality by analyzing code quality. Recovery is employed as a technique to check the consistency between SA and SC for improving the maintainability and reliability of systems as well as avoiding architectural erosion. Furthermore, there is a need for practitioners to adopt more dedicated approaches and tools to identify, analyze and use the relationships between SA and SC, especially SA recovery from SC. Whilst a few dedicated approaches and tools have been reported in the literature, the survey results show that practitioners feel that there are not many suitable tools to help identify and understand SA and SC relationships. Therefore, it is beneficial for practitioners to collaborate with researchers to have dedicated tools and approaches developed and adopted for continuously bridging the gap between SA and SC in the contemporary software development practices, e.g., agile development [7].

**For researchers** - our study results lead to the future research directions in the area of the relationships between SA and SC. For example, the study has identified five features of relationships and their frequencies; this information raises an important question about the types of relationships that may interest to practitioners the most. Moreover, the study has also provided an evidence of lack of suitable tools for which the future research efforts can be allocated to meet the needs of practitioners when identifying, analyzing and using these five features of relationships. A systematic classification of the relationships with the corresponding approaches and tools is needed through conducting a literature review, which can help practitioners to be aware of what and how the relationships between SA and SC can support various development activities (e.g., software testing) by identifying, analyzing and using the relationships in software development. More importantly, dedicated approaches and tools should be integrated with a development environment, such as Eclipse, to alleviate the cost and effort for bridging the gap between SA and SC. Finally, more empirical studies are needed to quantify the tangible benefits of managing the relationships and persuade practitioners to adopt dedicated approaches and tools for identifying, analyzing and using the relationships between SA and SC.

# 7 Threats to Validity

We discuss the threats to the validity of this study by following the guideline [54].

## 7.1 Internal Validity

Internal validity refers to variables that impact data analysis and results. One factor is the respondents' opinion and behavior could change in the short term. To deal with this threat, we limited the time of completing the questionnaire to 15 minutes, which was also verified in the pilot survey. Furthermore, the questionnaire can only be submitted with all questions finished, which makes sure that all the respondents completed the questionnaire in a similar period of time.

Another factor that could impact our survey results is that the respondents and interviewees did not have the suitable experience and expertise to answer the questions. To address this issue, we explicitly stated the characteristics of the target population in the survey letter and applied the inclusion and exclusion criteria for selecting valid samples. Moreover, to ensure the competence of the respondents to answer the questions, we only consider the responses with all the questions completed in English, because incomplete questions could suspiciously sabotage a study.

The last threat to the internal validity regards the no-response bias or error when a large number of the respondents demotivate to answer a survey, which could severely compromise the results of a survey. If only those participants with positive experience could likely spend time to complete a survey and provide valuable feedback, it could indicate self-selection bias. To alleviate the potential implications of this threat, we promised to share the survey results with all the respondents to increase their motivation of participation.

## 7.2 Construct Validity

Construct validity concerns the correctness and comprehensibility of the survey questionnaire and interview instrument. In this study, construct validity refers to whether or not the survey and interview questions can rationally support the results, because the questions could impact the survey results if badly designed. To mitigate this threat, we designed the precise questions in the questionnaire based on the relevant literature. All the questions in the questionnaire and interview were checked by the first two authors and any disagreements were discussed between these two authors. Especially the open-ended questions in the survey could be misinterpreted and filled in with irrelevant or confusing answers. We also invited the other two researchers from the research area to provide feedback on the understandability of the questions of this study. Based on the answers and feedback, we refined the questionnaire and interview instrument. Moreover, another threat is that a comprehensive answer for open-ended questions may not be obtained due to the time limitation within 15 minutes. To alleviate this threat, we adopted the following strategies: a pilot survey was conducted with five respondents which shows that the survey questionnaire can be completed with a comprehensive answer within 15 minutes; we also excluded the responses with invalid or insufficient answers to ensure the quality of the data (i.e., included survey responses); an additional interview with 8 practitioners was conducted to collect additional insights and more comprehensive answers from the practitioners. Although, this may pose another threat that the limited number of interviewees is not a representative sample of the practitioners, the answers from the 8 interviewees can, to some extent, provide interesting insights which complement the survey results.

Another threat arises from whether the participants provide truthful answers because they may act differently from their real thoughts. To minimize this threat and motivate participants to answer the survey questionnaire and participate the interview with their own practical experience, we confirmed in the invitation letter that the survey and interview are anonymous without disclosing any identities of respondents.

### 7.3 External Validity

External validity refers to the extent of generalizability of the findings. In this study, we intended to investigate the understanding of the relationships between SA and SC from the perspective of practitioners who have experience in architecture and implementation. The results of demographic data show that most participants work on small-to-medium-sized projects (< 250 developers), which may induce the risk of an unbalanced sample. Therefore, one main threat of this study is related to whether the 87 respondents and 8 interviewees are a good and strong representative of the software practitioners whose job includes the responsibility of dealing with SA and SC. We cannot eliminate this threat because we used a Non-Probabilistic Sampling method to recruit the participants for the online survey and interview, hence, the survey results cannot be considered statistically generalizable [12].

Another threat is about the response rate. We received 87 valid responses with 10.3% response rate, which is relatively small. There is no explicit standard or guideline for the acceptable response rate for software engineering survey studies except the self-reported data by some researchers. For example, Chakraborty *et al*. [55] and Bosu *et al*. [56] conducted their surveys with a response rate of ~13% and ~16% respectively. We carefully analyzed the participants' profiles in GitHub and were satisfied that the respondents represent a diverse background and suitably qualified for participating in this study. This makes us confident that the results are presentative and generalizable to some extent, as Kitchenham and Pfleeger suggested that "*low response rates may not imply lower representativeness and achieving higher response rates do not necessarily mean more accurate results*" [12].

### 7.4 Reliability

Conclusion validity concerns whether the survey is repeatable when other researchers conduct it. To mitigate the threats to reliability, we specified the process of survey design (i.e., creating the questionnaire, obtaining and analyzing the data) and interview design (i.e., creating interview instrument, participants interview, and transcript coding and analysis). One potential threat is the coding process of the qualitative data because it could have introduced personal bias. We conducted a pilot data analysis before the formal coding and analysis. After forming a consensus on the results of the pilot data analysis, the first author conducted the data coding and analysis using Grounded Theory; the other authors checked the results and the disagreements were discussed for reaching a consensus on the analysis. We have released the survey questionnaire, interview instrument, and all the data (87 survey responses, 8 interview transcripts and the encoded data) of this study [47] as well as the questionnaire link online [46]. The measure will make it easier for other researchers to verify the results of and replicate this study.

# 8 Conclusions

This paper has reported an empirical study aimed at investigating the practitioners' perception of the relationships between SA and SC. For this purpose, we conducted an online survey with 87 responses and 8 interviews with practitioners to collect and analyze data to understand how practitioners consider the relationships between SA and SC, the approaches and tools used for identifying, analyzing, and using the relationships, and the benefits and difficulties of identifying, analyzing, and using the relationships.

Our results reveal that: (1) The five features of relationships (Transformability, Traceability, Consistency, Interplay, and Recovery) between SA and SC reported in the literature are all considered and used by practitioners. (2) A few dedicated approaches and tools in the literature are employed by practitioners to identify, analyze, and use specific relationships, such as consistency. (3) The practitioners agreed that it is critical to identify, analyze, and use the relationships in order to improve system qualities, especially maintainability and reliability. (4) The cost and effort are the major barriers that hinder practitioners from identifying, analyzing, and using the relationships.

There are several future research directions according to the survey results: (1) To conduct a semi-structured interview to deeply analyze the difficulties as well as the specialized approaches of identifying and using the relationships between SA and SC. (2) To develop a systematic framework to manage the five features of relationships between SA and SC with dedicated approaches and tools support. (3) To systematically review how to quantify the cost and benefit of maintaining the relationships with approaches and tools from the literature.

# Acknowledgements

This work is partially sponsored by the National Key R&D Program of China with Grant No. 2018YFB1402800. The authors gratefully acknowledge the financial support from the China Scholarship Council.

# Appendix A. Survey Questionnaire

Table 12.    The welcome page of the survey

| |
|---|
| Note: If you have completed this questionnaire, please do not repeat!<br>The prerequisites for completing the questionnaire:<br>1. You have worked on software development.<br>2. You have worked on architecture design or software design.<br>This questionnaire aims to investigate the relationships between software architecture and source code, and how to identify, analyze, and use these relationships to facilitate software development. This questionnaire will take about 10-15 mins. The results of the survey will be shared with all the participants via the email addresses recorded.<br>* No personally identifiable information will be associated with your responses to any reports of these data. All information provided will be treated strictly as confidential and purely for academic purpose. |

Table 13.    Survey questions

| Type of Questions | Questions | Type of Answers |
|---|---|---|
| Background information about participants | Q1 Which country are you working in? | Free text |
| | Q2 What is your highest degree obtained in computer science or related fields? | BSc / MSc / PhD / Others |
| | Q3 (Multiple choices) What are your main role and function in company projects? | Project manager / Team leader / Architect / Developer / Consultant / Tester / Researcher / Others |
| | Q4 How long (in months) have you worked in software development? | Positive Number |
| | Q5 How long (in months) have you worked as a software architect or designer? | Positive Number |
| | Q6 What is the size of the largest project you participated in (number of developers)? | Population: <10 / 10-50 / 50-250 / >250 |
| | Q7 (Multiple choices) What are the domains or areas you worked in during your careers? | Embedded system / E-commerce / Financial / Healthcare / Telecommunication / Retail / Insurance / Other domains |
| | Q8 Your email address for receiving the results of the survey | Free text |
| Questions for answering RQ1 | Q9 Are there certain types of relationships between software architecture and code? | Yes / No |
| | Q10 Why do you think there are certain types of relationships between software architecture and code? | Free text |
| | Q11 If you know certain types of relationships between software architecture and | Free text |
| | code, please provide this (these) relationships? | |
| | Q12 Have you ever considered these relationships in your work? | Yes / No |
| | Q13 Which types of relationships do you frequently consider in your work? | Free text |
| | Q14 How important are the relationships between software architecture and code in software development? | 1-6 (1 = Unimportant, 2 = Of little importance, 3 = Moderately important, 4 = Important, 5 = Very important, 6 = No idea) |
| Questions for answering RQ2 | Q15 How did you identify and analyze the relationships between software architecture and code? If approaches or tools are available, can you describe and explain how to use them? | Free text |
| | Q16 How did you use the relationships between software architecture and code? If approaches or tools available, can you describe them and explain how to use them? | Free text |
| | Q17 What are the benefits of identifying, analyzing, and using the relationships between SA and code? | Free text |
| Questions for answering RQ3 | Q18 What are the limitations of identifying, analyzing, and using the relationships between SA and code? | Free text |

# Appendix B. Interview Instrument

Table 14. The welcome message for the interview

| |
|---|
| The prerequisites for participating the interview: <br> 1. You have worked on software development. <br> 2. You have worked on architecture design or software design. <br>     This interview aims to get deeper insights into the previous survey results about the relationships between software architecture and source code. This interview is expected to last about 30 to 45 minutes. The results of the interview will be shared with all the participants via the email addresses recorded. <br>     * No personally identifiable information will be associated with your responses to any reports of these data. All information provided will be treated strictly as confidential and purely for academic purpose. |

Table 15. Interview questions

| Type of Questions | Questions | Type of Answers |
|---|---|---|
| | IQ1.1 Which country are you working in? | Free text |
| Background information about participants | IQ1.2 (Multiple choices) What are your main role and function in company projects? | Project manager / Team leader / Architect / Developer / Consultant / Tester / Researcher / Others |
| | IQ1.3 How long (in months) have you worked in software development? | Positive Number |
| | IQ1.4 How long (in months) have you worked as a software architect or designer? | Positive Number |
| | IQ1.5 What is the size of the largest project you participated in (number of developers)? | Population: <10 / 10-50 / 50-250 / >250 |
| | IQ1.6 (Multiple choices) What are the domains or areas you worked in during your careers? | Embedded system / E-commerce / Financial / Healthcare / Telecommunication / Retail / Insurance / Other domains |
| Questions for answering RQ1 | Our survey results show that five features of relationships (Transformability, Traceability, Consistency, Influence, and Recovery) between software architecture and source code are known and considered by practitioners (we will introduce these five relationship features briefly to the participants). We want to know: | |
| | IQ2.1.1 Whether and how do you transform software architecture to source code? | Free text |
| | IQ2.1.2 Whether and how do you build and maintain the trace links between software architecture and source code? | Free text |
| | IQ2.1.3 Whether and how do you ensure the consistency between software architecture and source code? | Free text |
| | IQ2.1.4 Whether and how can the quality of software architecture influence the quality of source code? | Free text |
| | IQ2.1.5 Whether and how can the quality of source code influence the quality of software architecture? | Free text |
| | IQ2.1.6 Whether and how do you recover software architecture from source code? | Free text |
| | IQ2.1.7 Which types of relationships do you frequently consider and use in your work? Please give a reason(s) to support your answer. | Free text |
| | IQ2.1.8 Do you think there are interrelationships between the five features of relationships (Yes OR No)? Please give some examples with a reason(s) to support your answer. | Free text |
| | IQ2.2.1 What architectural concepts and knowledge (e.g., architecture views) are considered important to support the relationships between software architecture and source code? | Free text |
| | IQ2.2.2 How can architectural concepts and knowledge (e.g., architecture views) be used to support the relationships between software architecture and source code? | Free text |
| | IQ2.2.3 What is the interplay between architecture decisions and implementation decisions? | Free text |
| | Our survey results show that the practitioners only mentioned static traceability between structural architecture elements and code elements without further considering dynamic traceability at runtime between source code and software architecture. | |
| | IQ2.3 Do you consider and use dynamic traceability between software architecture and source code (Yes OR No)? Please give a reason(s) to support your answer. | Free text |
| Questions for answering RQ2 | Our survey results show that the type of approaches adopted by most respondents for identifying, analyzing, and using the relationships between software architecture and source code are the top-down integration approaches, such as Pattern-Oriented Design (POD), Domain-Driven Design (DDD), and Service-Oriented Development (SOD). | |
| | IQ3.1 Which approach do you often adopt to identify, use, and analyze the relationships between software architecture and code? Please give a reason(s) to explain your answer. | Free text |
| | Our survey results show that dedicated approaches and tools proposed in the literature are less adopted by practitioners to identify, analyze, and use the relationships between software architecture and source code. | |
| | IQ3.2 Do you ever use any dedicated approaches and tools to identify, use, and analyze the relationships between software architecture and code (Yes OR No)? Please give some approaches and tools with a reason(s) to support your answer. | Free text |
| Questions for answering RQ3 | Our survey results show that most of the respondents realized that quality attributes (especially maintainability and reliability) of systems are improved through identifying, analyzing, and using the relationships between software architecture and source code. We want to know: | |
| | IQ4.1 How can the quality attributes (especially maintainability and reliability) of systems be improved through identifying, analyzing, and using the relationships between software architecture and source code? | Free text |
| | Our survey results show that the main barrier for practitioners to identify, analyze, and use the relationships between software architecture and source code in practice is the cost and effort needed for bridging their gap. We want to know: | Free text |
| | IQ4.2 How do you analyze the costs (effort) and benefits when identifying, analyzing, and using the relationships between software architecture and source code? | Free text |

# References

[1] Kruchten, P., Obbink, H., and Stafford, J., 2006. The past, present, and future for software architecture. IEEE Software, 23(2): 22-30.
[2] Perry, D.E. and Wolf, A.L., 1992. Foundations for the study of software architecture. ACM SIGSOFT Software Engineering Notes, 17(4): 40-52.
[3] Medvidovic, N. and Taylor, R.N., 2000. A classification and comparison framework for software architecture description languages. IEEE Transactions on Software Engineering, 26(1): 70-93.
[4] Reja, U., Manfreda, K.L., Hlebec, V. and Vehovar, V., 2003. Open-ended vs. close-ended questions in web questionnaires. Developments in Applied Statistics, 19(1): 159-177
[5] Jansen, A. and Bosch, J., 2005. Software architecture as a set of architectural design decisions. In: Proceeding of the 5th Working IEEE/IFIP Conference on Software Architecture (WICSA). Pittsburgh, PA, USA, pp. 109-120.
[6] Bass, L., Clements, P., and Kazman, R., 2012. Software Architecture in Practice, 3rd Edition, Addison-Wesley Professional.
[7] Murta, L.G., van der Hoek, A., and Werner, C.M., 2008. Continuous and automated evolution of architecture-to-implementation traceability links. Automated Software Engineering, 15(1): 75-107.
[8] Macia, I., Arcoverde, R., Garcia, A., Chavez, C., and von Staa, A., 2012. On the relevance of code anomalies for identifying architecture degradation symptoms. In: Proceedings of the 16th European Conference on Software Maintenance and Reengineering (CSMR). Szeged, Hungary, pp. 277-286.
[9] Brown, S., 2017. Software Architecture for Developers - Volume 2: Visualise, Document and Explore Your Ssoftware Architecture. Lean Publishing.
[10] Fairbanks, G., 2010. Just Enough Software Architecture: A Risk-Driven Approach. Marshall & Brainerd.
[11] Shaw, M. and Clements, P., 2006. The golden age of software architecture. IEEE Software, 23(2): 31-39.
[12] Kitchenham, B.A. and Pfleeger, S.L., 2008. Personal Opinion Surveys, in Guide to Advanced Empirical Software Engineering. Springer London, pp. 63-92.
[13] Fink, A., 2003. The Survey Handbook. 2nd Edition. Sage Publications.
[14] Fowler Jr, F.J., 2013. Survey Research Methods. 5th Edition. Sage publications.
[15] Javed, M.A. and Zdun, U., 2014. A systematic literature review of traceability approaches between software architecture and source code. In: Proceedings of the 18th International Conference on Evaluation and Assessment in Software Engineering (EASE). London, England, United Kingdom, Article No. 16.
[16] Buchgeher, G. and Weinreich, R., 2011. Automatic tracing of decisions to architecture and implementation. In: Proceedings of the 9th Working IEEE/IFIP Conference on Software Architecture (WICSA). Washington, DC, pp. 46-55.
[17] Stevanetic, S., Haitzer, T., and Zdun, U., 2014. Supporting software evolution by integrating DSL-based architectural abstraction and understandability related metrics. In: Proceedings of the 8th ACM European Conference on Software Architecture Workshops (ECSAW). Vienna, Austria, Article No. 19.
[18] Nguyen, T. N., Munson, E. V., and Thao, C., 2005. Object-oriented configuration management technology can improve software architectural traceability. In: Proceedings of the 3rd ACIS International Conference on Software Engineering Research, Management and Applications (SERA). Washington, DC, USA, pp. 86-93.
[19] Santos, J.C.S., Mirakhorli, M., Mujhid, I., and Zogaan, W., 2016. BUDGET: A tool for supporting software architecture traceability research, In: Proceedings of the 13th IEEE/IEIP Conference on Software Architecture (WICSA). Venice, Italy, pp. 303-306.

[20] Zheng, Y. and Taylor, R.N., 2012. xMapper: An architecture-implementation mapping tool. In: Proceedings of the 34th International Conference on Software Engineering (ICSE). Zurich, Switzerland, pp. 1461-1462.

[21] Ali, N., Baker, S., O'Crowley, R., Herold, S., and Buckley, J., 2018. Architecture consistency: state of the practice, challenges and requirements. Empirical Software Engineering, 23(1): 224-258.

[22] Buckley, J., Mooney, S., Rosik, J., and Ali, N., 2013. JITTAC: a just-in-time tool for architectural consistency. In: Proceedings of the 35th International Conference on Software Engineering (ICSE). San Francisco, CA, USA, pp. 1291-1294.

[23] De Silva, L. and Balasubramaniam, D., 2012. Controlling software architecture erosion: A survey. Journal of Systems and Software, 85(1): 132-151.

[24] Herold, S., Blom, M., and Buckley, J., 2016. Evidence in architecture degradation and consistency checking research: preliminary results from a literature review. In: Proceedings of the 10th European Conference on Software Architecture Workshops (ECSAW). Copenhagen, Denmark, Article No. 20.

[25] Muccini, H., Dias, M., and Richardson, D.J., 2004. Systematic Testing of Software Architectures in the C2 style. In: Proceedings of International Conference on Fundamental Approaches to Software Engineering (FASE). Springer, Berlin, Heidelberg, pp. 295-309.

[26] Muccini, H., Inverardi, P., and Bertolino, A., 2004. Using software architecture for code testing. IEEE Transactions on Software Engineering, 30(3): 160-171.

[27] Passos, LT., Terra, R., Diniz, M., Valente, R., and Mendonça, NC., 2010. Static architecture-conformance checking: an illustrative overview. IEEE Software, 27(5): 82-89.

[28] Murphy, G.C., Notkin, D., and Sullivan, K.J., 1995. Software reflexion models: bridging the gap between source and high-level models. In: Proceeding of the 3rd ACM SIGSOFT Symposium on Foundations of Software Engineering (FSE), Washington, D.C., USA. pp 18-23.

[29] Ali, N., Rosik, J., and Buckley, J., 2012. Characterizing real-time reflexion-based architecture recovery: an in-vivo multicase study. In: Proceedings of the 8th International ACM SIGSOFT Conference on Quality of Software Architectures (QoSA). New York, USA, pp. 23-32.

[30] Rosik, J., Buckley, J., and Babar, M.A., 2009. Design Requirements for an Architecture Consistency Tool. In: Proceedings of the 21st Working Conference of the Psychology of Programmers' Interest Group (PPIG), pp. 109-124.

[31] Brunet, J., Murphy, G.C., Serey, D., and Figueiredo, J., 2014. Five years of software architecture checking: A case study of Eclipse. IEEE Software, 32(5): 30-36.

[32] Buckley, J., Ali, N., English, M., Rosik, J., and Herold, S., 2015. Real-Time Reflexion Modelling in architecture reconciliation: A multi case study. Information and Software Technology, 61: 107-123.

[33] Medvidovic, N., Egyed, A., and Gruenbacher, P., 2003. Stemming architectural erosion by coupling architectural discovery and recovery. In: Proceeding of the 2nd International Software Requirements to Architectures Workshop (STRAW), Portland, Oregon, USA, pp. 61-68.

[34] Garcia, J., Ivkovic, I., and Medvidovic, N., 2013. A comparative analysis of software architecture recovery techniques. In: Proceedings of the 28th IEEE/ACM International Conference on Automated Software Engineering (ASE), Silicon Valley, CA, USA, pp. 486-496.

[35] Shahbazian, A., Lee, Y.K., Le, D., Brun, Y., and Medvidovic, N., 2018. Recovering architectural design decisions. In: Proceedings of 2018 IEEE International Conference on Software Architecture (ICSA). Seattle, WA, USA, pp. 95-104.

[36] Mirakhorli, M. and Cleland-Huang, J., 2016. Detecting, tracing, and monitoring architectural tactics in code. IEEE Transactions on Software Engineering, 42(3): 205-220.

[37] Turhan, N.K. and Oğuztüzün, H., N., 2016. Metamodeling of reference software architecture and automatic code generation. In: Proceedings of the 10th European Conference on Software Architecture Workshops (ECSAW). Copenhagen, Denmark, Article No. 2.

[38] Aldrich, J., Chambers, C., and Notkin, D., 2002. ArchJava: connecting software architecture to implementation. In: Proceedings of the 24th International Conference on Software Engineering (ICSE). Orlando, FL, USA, pp. 187-197.

[39] Gui, S., Ma, L., Luo, L., Yin, L., and Li, Y., 2008. UCaG: an automatic C code generator for AADL based upon DeltaOS. In: Proceedings of the 2008 International Conference on Advanced Computer Theory and Engineering (ICACTE). Phuket, Thailand, pp. 346-350.

[40] Pelliccione, P., Inverardi, P., and Muccini, H., 2008. Charmy: a framework for designing and verifying architectural specifications. IEEE Transactions on Software Engineering, 35(3): 325-346.

[41] Woods, E. and Rozanski, N., 2010. Unifying software architecture with its implementation. In Proceedings of the Fourth European Conference on Software Architecture: Companion Volume (ICSA). Copenhagen, Denmark, pp. 55-58.

[42] Fontana, F.A., Ferme, V., and Zanoni, M., 2015. Towards assessing software architecture quality by exploiting code smell relations. In: Proceedings of the 2nd International Workshop on Software Architecture and Metrics (SAM). Florence, Italy, pp. 1-7.

[43] Pigazzini, I., 2019. Automatic detection of architectural bad smells through semantic representation of code. In: Proceedings of the 13th European Conference on Software Architecture (ECSA), Paris, France, pp. 59-62.

[44] Vidal, S., Oizumi, W., Garcia, A., Pace, A.D., and Marcos, C., 2019. Ranking architecturally critical agglomerations of code smells. Science of Computer Programming, 182: 64-85.

[45] Santos, J.C., Tarrit, K., Sejfia, A., Mirakhorli, M., and Galster, M., 2019. An empirical study of tactical vulnerabilities. Journal of Systems and Software, 149: 263-284.

[46] A Survey on the Relationships between Software Architecture and Source Code: Online Survey Questionnaire. https://goo.gl/forms/hS4XVptbbiMAyUai1

[47] Tian, F., Liang, P., and Babar, M.A., 2021. Complementary Material for the Paper: "Relationships between Software Architecture and Source Code in Practice: An Exploratory Survey and Interview". http://doi.org/10.5281/zenodo.3889140

[48] Wohlin, C., Höst, M., and Henningsson, K., 2003. Empirical research methods in software engineering. In Empirical Methods and Studies in Software Engineering. Springer, Berlin, Heidelberg. pp. 7-23.

[49] Glaser, B.G. and Strauss, A.L., 2009. The Discovery of Grounded Theory: Strategies for Qualitative Research. Transaction Publishers.

[50] Adolph, S., Hall, W., and Kruchten, P., 2011. Using grounded theory to study the experience of software development. Empirical Software Engineering, 16(4): 487-513.

[51] Keim, J. and Koziolek, A., 2019. Towards consistency checking between software architecture and informal documentation. In: Proceedings of IEEE International Conference on Software Architecture Companion (ICSA-C), Hamburg, Germany, pp. 250-253.

[52] Falessi, D., Cantone, G., Sarcia, S.A., Calavaro, G., Subiaco, P., and D'Amore, C., 2010. Peaceful coexistence: Agile developer perspectives on software architecture. IEEE Software, 27(2): 23-25.

[53] ISO/IEC, ISO/IEC 25010:2011, Systems and software engineering - Systems and software Quality Requirements and Evaluation (SQuaRE) - System and software quality models.

[54] Wohlin, C., Runeson, P., Höst, M., Ohlsson, M.C., Regnell, B., and Wesslén, A., 2012. Experimentation in Software Engineering. Springer Science & Business Media.

[55] Chakraborty, P., Shahriyar, R., Iqbal, A., and Bosu, A., 2018. Understanding the software development practices of blockchain projects: a survey. In: Proceedings of the 12th ACM/IEEE International Symposium on Empirical Software Engineering and Measurement (ESEM), Oulu, Finland, Article No. 28.

[56] Bosu, A., Carver, J.C., Bird, C., Orbeck, J., and Chockley, C., 2016. Process aspects and social dynamics of contemporary code review: Insights from open source development and industrial practice at microsoft. IEEE Transactions on Software Engineering, 43(1): 56-75.

[57] Díaz-Pace, J.A., Soria, Á., Rodríguez, G., and Campo, M.R., 2012. Assisting conformance checks between architectural scenarios and implementation. Information and Software Technology, 54(5): 448-466.

[58] Eden, A.H. and Kazman, R., 2003. Architecture, design, implementation. In: Proceedings of the 25th International Conference on Software Engineering (ICSE), Portland, OR, USA, pp. 149-159.

[59] Le Gear, A., Buckley, J., Collins, J.J. and O'Dea, K., 2005. Software reconnexion: understanding software using a variation on software reconnaissance and reflexion modelling. In: Proceedings of the International Symposium on Empirical Software Engineering (ESE), Noosa Heads, QLD, Australia, pp. 10-pp.

[60] Wohlrab, R., Eliasson, U., Pelliccione, P. and Heldal, R., 2019. Improving the consistency and usefulness of architecture descriptions: Guidelines for architects. In: Proceedings of the 2019 IEEE International Conference on Software Architecture (ICSA), Hamburg, Germany, pp. 151-160.

[61] Zahid, M., Mehmmod, Z. and Inayat, I., 2017. Evolution in software architecture recovery techniques - A survey. In: Proceedings of the 13th International Conference on Emerging Technologies (ICET), Islamabad, Pakistan, pp. 1-6.

[62] de Silva, L. and Balasubramaniam, D., 2013. PANDArch: A pluggable automated non-intrusive dynamic architecture conformance checker. In: Proceedings of the European Conference on Software Architecture (ECSA), Berlin, Heidelberg, pp. 240-248.

[63] Ganesan, D., Keuler, T. and Nishimura, Y., 2008. Architecture compliance checking at runtime: an industry experience report. In: Proceedings of the 8th IEEE International Conference on Quality Software (QSIC), Oxford, UK, pp. 347-356.

[64] Lenhard, J., Blom, M. and Herold, S., 2019. Exploring the suitability of source code metrics for indicating architectural inconsistencies. Software Quality Journal, 27(1): 241-274.

[65] Martini, A. and Bosch, J., 2015. The danger of architectural technical debt: Contagious debt and vicious circles. In: Proceedings of the 12th Working IEEE/IFIP Conference on Software Architecture (ICSA), Montreal, QC, Canada, pp. 1-10.

[66] Kazman, R., Klein, M. and Clements, P., 2000. ATAM: Method for architecture evaluation. Technical Report CMU/SEI-2000-TR-004. Pittsburg: Carnegie Mellon University, Software Engineering Institute.

[67] The Open Group, 2018. The TOGAF Standard, Version 9.2. https://www.opengroup.org/togaf-standard-version-92-overview